\documentclass[twocolumn]{aastex631}
\usepackage{amsmath}
\usepackage{graphicx}
\usepackage[caption=false]{subfig}
\usepackage{float}
\usepackage{multirow,array,booktabs}
\usepackage{bm}

\newcommand{\Msol}{$M_\odot$}
\DeclareUnicodeCharacter{2212}{-}

\newcommand{\caltech}{Department of Astronomy, California Institute of Technology, Pasadena, CA 91125, USA}
\newcommand{\gps}{Division of Geological \& Planetary Sciences, California Institute of Technology, Pasadena, CA 91125, USA}
\newcommand{\ucsc}{Department of Astronomy \& Astrophysics, University of California, Santa Cruz, CA95064, USA}
\newcommand{\keck}{W. M. Keck Observatory, 65-1120 Mamalahoa Hwy, Kamuela, HI, USA}
\newcommand{\ucla}{Department of Physics \& Astronomy, 430 Portola Plaza, University of California, Los Angeles, CA 90095, USA}
\newcommand{\jpl}{Jet Propulsion Laboratory, California Institute of Technology, 4800 Oak Grove Dr.,Pasadena, CA 91109, USA}
\newcommand{\ucsdp}{Department of Physics, University of California, San Diego, La Jolla, CA 92093}
\newcommand{\ucsda}{Department of Astronomy and Astrophysics, University of California, San Diego, La Jolla, CA 92093}

\newcommand{\osu}{Department of Astronomy, The Ohio State University, 100 W 18th Ave, Columbus, OH 43210 USA}

\newcommand{\arizona}{James C. Wyant College of Optical Sciences, University of Arizona, Meinel Building 1630 E. University Blvd., Tucson, AZ. 85721}

\begin{document}
\title{Orbital and Atmospheric Characterization of the 1RXS J034231.8+121622 System Using High-Resolution Spectroscopy Confirms That The Companion is a Low-Mass Star}
\author[0000-0001-5173-2947]{Clarissa R. Do Ó}
\affiliation{\ucsdp}

\author[0000-0003-1399-3593]{Ben Sappey}
\affiliation{\ucsdp}

\author[0000-0002-9936-6285]{Quinn M. Konopacky}
\affiliation{\ucsda}

\author[0000-0003-2233-4821]{Jean-Baptiste Ruffio}
\affiliation{\ucsda}
\author[0000-0003-2400-7322]{Kelly K. O'Neil}
 \affiliation{Department of Physics, University of Nevada, Reno, Reno, NV 89557}
\author[0000-0001-9554-6062]{Tuan Do}
 \affiliation{\ucla}
 \author{Gregory Martinez}
 \affiliation{\ucla}
 \author[0000-0002-7129-3002]{Travis S. Barman}
 \affiliation{Lunar and Planetary Laboratory, University of Arizona, Tucson, AZ 85721, USA}
\author[0000-0002-9242-9052]{Jayke S. Nguyen}
\affiliation{\ucsda}

\author[0000-0002-6618-1137]{Jerry W. Xuan}
 \affiliation{\caltech}
 \author[0000-0002-9807-5435]{Christopher A. Theissen}
 \affiliation{\ucsda}
\author[0000-0002-3199-2888]{Sarah Blunt}
\affiliation{Center for Interdisciplinary Exploration and Research in Astrophysics (CIERA) and Department of Physics and Astronomy, Northwestern University, Evanston, IL 60208, USA}
\author[0000-0001-5684-4593]{William Thompson}
\affiliation{
   National Research Council of Canada Herzberg,
5071 West Saanich Rd,
Victoria, BC, V9E 2E7, Canada}

\author[0000-0002-5370-7494]{Chih-Chun Hsu}
\affiliation{Center for Interdisciplinary Exploration and Research in Astrophysics (CIERA) and Department of Physics and Astronomy, Northwestern University, Evanston, IL 60208, USA}

\author{Ashley Baker}
 \affiliation{\caltech}
 
\author{Randall Bartos}
\affiliation{\jpl}

\author{Geoffrey A. Blake}
\affiliation{\gps}

\author[0000-0003-4737-5486]{Benjamin Calvin}
\affiliation{\ucla}
\affiliation{\caltech}

\author{Sylvain Cetre}
\affiliation{\keck}

% Delorme
\author[0000-0001-8953-1008]{Jacques-Robert Delorme}
\affiliation{\keck}

% Doppman
\author{Greg Doppmann}
\affiliation{\keck}

\author{Daniel Echeverri}
\affiliation{\caltech}

\author[0000-0002-1392-0768]{Luke Finnerty}
\affiliation{\ucla}

% Fitzgerald
\author[0000-0002-0176-8973]{Michael P. Fitzgerald}
\affiliation{\ucla}

\author[0000-0001-9164-7966]{Julie Inglis}
\affiliation{\gps}

% Jovanovic
\author[0000-0001-5213-6207]{Nemanja Jovanovic}
\affiliation{\caltech}

% Lopez
\author[0000-0002-2019-4995]{Ronald A. L\'opez}
\affiliation{\ucla}

\author{Dimitri Mawet}
\affiliation{\caltech} 
\affiliation{\jpl}

% Morris
\author{Evan Morris}
\affiliation{\ucsc}

% Pezzato
\author{Jacklyn Pezzato}
\affiliation{\caltech}

% Schofield
\author{Tobias Schofield}
\affiliation{\caltech}

% Skemer
\author{Andrew Skemer}
\affiliation{\ucsc}

% Wallace
\author[0000-0001-5299-6899]{J. Kent Wallace}
\affiliation{\jpl}

\author[0000-0003-0774-6502]{Jason J. Wang}
% \affiliation{\caltech}
\affiliation{Center for Interdisciplinary Exploration and Research in Astrophysics (CIERA) and Department of Physics and Astronomy, Northwestern University, Evanston, IL 60208, USA}

% Wang (王吉)
\author[0000-0002-4361-8885]{Ji Wang}
\affiliation{\osu}

\author[0000-0002-4934-3042]{Joshua Liberman}
\affiliation{\arizona}

\begin{abstract}
The 1RXS J034231.8+121622 system consists of an M dwarf primary and a directly imaged low-mass stellar companion. We use high resolution spectroscopic data from Keck/KPIC to estimate the objects’ atmospheric parameters and radial velocities (RVs). Using PHOENIX stellar models, we find that the primary has a temperature of 3460 $\pm$ 50 K a metallicity of 0.16 $\pm$ 0.04, while the secondary has a temperature of 2510 $\pm$ 50 K and a metallicity of $0.13\substack{+0.12 \\ -0.11}$. Recent work suggests this system is associated with the Hyades, placing it an older age than previous estimates. Both metallicities agree with current $[Fe/H]$ Hyades measurements (0.11 -- 0.21). Using stellar evolutionary models, we obtain significantly higher masses for the objects, of 0.30 $\pm$ 0.15 $M_\odot$ and 0.08 $\pm$ 0.01 $M_\odot$ (84 $\pm$ 11 $M_{Jup}$) respectively. Using the RVs and a new astrometry point from Keck/NIRC2, we find that the system is likely an edge-on, moderately eccentric ($0.41\substack{+0.27 \\ -0.08}$) configuration. We also estimate the C/O ratio of both objects using custom grid models, obtaining 0.42 $\pm$ 0.10 (primary) and 0.55 $\pm$ 0.10 (companion). From these results, we confirm that this system most likely went through a binary star formation process in the Hyades. The significant changes in this system’s parameters since its discovery highlight the importance of high resolution spectroscopy for both orbital and atmospheric characterization of directly imaged companions.
\end{abstract}

\section{Introduction} \label{intro}

Directly imaging substellar companions allows for the observation of the companion's thermally emitted light. This method of detection allows for their orbital and atmospheric characterization, providing valuable information on how these objects formed. The observation of these objects using ground-based telescopes such as the W. M. Keck Observatory requires an adaptive optics system (AO) coupled with an imaging instrument to obtain relative companion astrometry for orbital characterization and with a medium- or high-resolution spectrograph to obtain relative radial velocity (RVs) for further orbital characterization and spectral information on the companion. 
\par
Temperature, $\log(g)$, rotation speed, bulk metallicity, and elemental abundances that may be tracers of formation processes in these systems can be gleaned from spectra of the companion's and host star. Temperature, $log(g)$ and luminosity values coupled with a system's age are useful for inferring the mass of an object using evolutionary models (e.g. \citealt{Baraffe1998}, \citealt{Baraffe2003}, \citealt{Baraffe_2015}, \citealt{Mukherjee2024}, \citealt{Morley2024}). 
\par
Beyond these quantities, high resolution spectroscopy (R $>$ 20,000) can offer a means to measure elemental abundances, such as carbon and oxygen, and chemical ratios of these abundances by resolving individual absorption lines from different molecules containing these elements. All of these parameters can inform how these directly imaged companions have formed and whether they are exoplanets, brown dwarfs, or low mass stars. Therefore, directly imaging both high and low mass companions are important for better constraining and informing companion formation trends.
\par
Obtaining high resolution spectra is often challenging for directly imaged companions due to their faint magnitudes and close separations from their host stars. However, the Keck Planet Imager and Characterizer (KPIC) (\citealt{Delorme2021}) is an instrument designed for this task. KPIC is a series of upgrades to the Keck II Telescope's AO system and the NIRSPEC spectrograph, which includes a single-mode fiber injection unit coupled to the latter \citep[e.g.][]{McLean2000, Martin2018, Lopez2020}, allowing for the high resolution spectroscopy (R $\sim$ 35,000) of directly imaged companions in $K$ band \citep{Wang_2021, Wang2022, Xuan2022, Xuan2024}.
\subsection{Atmospheric Abundances}
The C/O ratio has been considered as a diagnostic of formation pathways for companions. If the companion went through a binary star formation process or protostellar disk instability process, it should present elemental abundances similar to that of the host star (\citealt{HelledSchubert2009}). For core or pebble accretion scenarios, however, the planet's assembly can generate a variety of C/O ratios, mostly dependent on the planet's location during its forming stage compared to the snow lines of volatile species such as CO, CO$_2$ and H$_2$O (\citealt{Oberg_2011}). Most recently, \citealt{Hoch_2023} assessed the C/O ratio as a formation tracer for several companions and found that some observed C/O ratio trends cannot be fully explained by planet formation models in simulations. They conclude that measuring the C/O ratio for a companion is a useful proxy for trends in companion formation pathways, but it is not an assertive value that completely constrains formation mechanisms in individual objects. Therefore, measuring the C/O ratio using high resolution spectroscopy can potentially inform companion formation trends, but may need to be considered in conjunction with other formation indicators. \par
Metallicity measurements may also be a useful proxy for the formation of companions. The metallicity $[Fe/H]$ of a companion is most likely enhanced compared to its host star if it presents a planet-like formation (e.g. core or pebble accretion; \citealt{HelledSchubert2009}), while metal enrichment is not expected if the companion formed from disk instability or disk fragmentation (e.g. \citealt{Boss2010}). Beyond the formation information obtained from this parameter, the metallicity is also useful for informing potential membership of specific objects with a cluster (e.g. \citealt{Perryman1998}), where objects in the same cluster often present similar metallicity values due to their shared formation history. \par 

\subsection{Rotational Velocity}
The rotational velocity can help constrain the angular momentum history of an object (e.g. \citealt{Bryan_2020}, \citealt{Bryan_2021}). Rotation rates of stars have been used to trace stellar activity (e.g. \citealt{Browning_2008}). For low mass stars and brown dwarfs specifically, it appears that these objects are common rapid rotators, with a dependence on the object's temperature, age and mass (e.g. \citealt{Mohanty2003}, \citealt{Zapa2006} \citealt{Reiners2008}, \citealt{Konopacky2012}, \citealt{Snellen2014}, \citealt{Bryan_2020}, \citealt{Tannock2021}, \citealt{Hsu_2021b}, \citealt{Wang_2021}, 
\citealt{Zendel2023}, \citealt{Landman2024}, Hsu et al. in prep). For younger objects ($\approx$ 10 -- 100s of Myr), the rapid rotation can be explained by the gravitational contraction of these objects, which causes them to spin up as they age due to the conservation of angular momentum. This contraction ends at $\approx$ 1 Gyr. The rotational velocity of these objects is also mass dependent (e.g. \citealt{Herbst2007}), with higher mass stars presenting longer rotation periods than brown dwarfs and low mass stars. \citealt{Reiners2008} found that lower mass stars and brown dwarfs tend to rotate faster because angular momentum loss mechanisms, such as magnetic braking, have longer timescales at lower mass and are therefore more inefficient at slowing down the objects’ spin. These authors therefore find that young low mass objects initially have intermediate rotation rates and accelerate due to contraction in the first few 10 Myr. After this timescale, the objects with M $>$ 0.09 \Msol begin to decelerate until they lose most of their angular momentum at 10s of Myr.  For low mass objects with M $<$ 0.09 \Msol, the braking law becomes so weak that even at older ages the objects spin faster.
\par
High resolution spectroscopy allows for the measurement of spin, or $v\sin(i)$, of a target. Eliminating the $\sin(i)$ degeneracy that would allow for a true rotational velocity measurement requires photometrically derived rotational periods (e.g. \citealt{Bowler_2023}). Several systems have been found to have primary and secondary rotation axis misaligned from the orbital plane (e.g. \citealt{Bowler_2023}, \citealt{Bryan_2020}, \citealt{Bryan_2021}), which could have a variety of causes, such as the direct consequence of formation in the protoplanetary disk (e.g. \citealt{Epstein-Martin_2022}), nearby mass concentrations during the collapse of the cloud core (e.g. \citealt{Tremaine_1991}) or interactions with objects outside of the forming system (e.g. \citealt{Anderson_2017}).  \par

\subsection{Orbital Parameters}
Orbital characterization of a companion provides dynamical footprints on its formation history, including possible distinctions between planet formation, such as core accretion and protoplanetary disk instability, and binary star formation, such as protostellar disk fragmentation (e.g. \citealt{Offner2022, Tokovinin2020}). One particularly important orbital parameter for tracing the formation history of an object is its orbital eccentricity (e.g. \citealt{Kipping_2013}, \citealt{Bowler_2020}, \citealt{Nagpal2023}, \citealt{DoO_2023}). In the theory of planet formation, companions that form via core or pebble accretion most likely present lower eccentricities (i.e., closer to circular orbits) due to the collision of particles in the protoplanetary disk, while formation via gravitational instability or binary star mechanisms may present higher eccentricities during formation (e.g. \citealt{Mayer_2004}).  Dynamical interactions after formation may also modify the initial eccentricities of planets, with the possibility of scattering planets to wide separations on high eccentricity orbits (e.g., \citealt{Chatterjee2008,Veras2009}). 
\par
Due to the large separation from the host stars and consequently long period of these objects, orbit characterization using relative astrometry of the companion is often limited. Generally, orbit fits are carried out in a Bayesian framework, where a time series of relative astrometry coupled with uniform parameter priors generates posteriors on the orbital parameters of the companion. Poorly sampled data of these orbits coupled with priors has been shown to have biases on the final parameters of the companion, in particular the eccentricity, which is an important dynamical tracer of these objects (\citealt{Greg2017}). This undersampling coupled with biased posteriors led to the generation of new prior approaches for orbit fitting that aimed to mitigate these biases, such as observable-based priors (\citealt{ONeil_2019}). However, for any priors, a single astrometry point for an undersampled orbit can significantly change the posterior distributions of orbital parameters. \par
Adding third dimensional information using relative radial velocity data between the companion and primary star to an orbit fit allows for a more constraining orbital characterization \citep[e.g.][]{Schwarz2016, Ruffio2019, DoO_2023, Xuan2024}. In particular, the radial velocity generally provides better constraints on the orbital plane of a companion due to the elimination of degeneracies in the angle of ascending node ($\Omega$) and the argument of periapsis angle ($\omega$) as well as eccentricity-inclination degeneracies. 

\subsection{The 1RXS J034231.8+121622 System}
The companion to the star 1RXS J034231.8+121622, or 2MASS J03423180+1216225, was initially a resolved companion candidate reported by \citealt{Janson_2012}. The data obtained by their work could not distinguish the candidate from a background object. The companion was later confirmed by \citealt{Bowler_2015a}, who found that the object was physically bound to the primary. The primary is a M4 type star which has shown signs of youth due to its X-ray emission (\citealt{Shkolnik2009}). The companion was initially classified as a L0 dwarf by the discovery paper, with a mass range of 35 $\pm$ 8 $M_\mathrm{Jup}$ using evolutionary models for the companion's luminosity, the assumed system age of 60 -- 300 Myr and system distance of 32.995 $\pm$ 0.0727 parsecs from Gaia DR2 (\citealt{GDR2}). However, \citealt{Kuzuhara_2022} recently found that this system is most likely associated with the Hyades, which has a much older age. They find a mass of 76 -- 83 $M_{Jup}$ for the companion using this new age. This result prompts a re-calculation of the primary mass and consequently overall system mass.
\par
In this work, we characterize the  1RXS J034231.8+121622 system. In order to better constrain the companion's orbit, we use Keck/NIRC2 data to directly image and obtain a new astrometric data point and Keck/KPIC data to obtain radial velocities and characterize the atmosphere (effective temperature $T_{\rm eff}$, surface gravity $log(g)$, metallicity $[Fe/H]$, spin $vsin(i)$ and C/O ratio) of the companion and the host star. \par

Using the information provided by high resolution spectroscopy, we constrain the possible formation scenarios for this specific companion. This work is organized as follows: in Section \ref{methods}, we present our data reduction to obtain the NIRC2 astrometry point (Section \ref{astrom}) and to reduce the KPIC spectrum (Section \ref{kpic}). Our results are presented in Section \ref{results}, with the temperature, $log(g)$ and $[Fe/H]$ fits shown in Section \ref{templogg}. The new system mass is derived in Section \ref{mass_section}, and the new orbital parameters are presented in Section \ref{masssys}. Section \ref{atmos} presents the C/O ratio analysis (Section \ref{CO}) and rotational velocities (Section \ref{rotations}) for both objects. We discuss and conclude our results in Sections \ref{disc} and \ref{conc} respectively.

\section{Methods} \label{methods}

\subsection{Keck/NIRC2 Data Reduction} \label{astrom}

We obtained astrometry for the 1RXS J034231.8+121622 system with the NIRC2 camera on the W.M. Keck Telescope II on UT 2023 February 8 using natural guide star adaptive optics with the primary as the guide star (\citealt{Wizinowich_2013}). This epoch of observation comes 5 years after the last published astrometry point from 2018 by \citet{Bowler_2020}. We take 30 exposures with exposure time of 0.5 seconds and 120 coadds. Our images do not use a coronagraph as the target is at $\approx$ 700 mas from the host star with a contrast of $3.2 * 10^{-2}$ and is easily visible in the raw frames.\par
\begin{figure}[htb!]
  \begin{center}
\centerline{\includegraphics[width=5.1in]{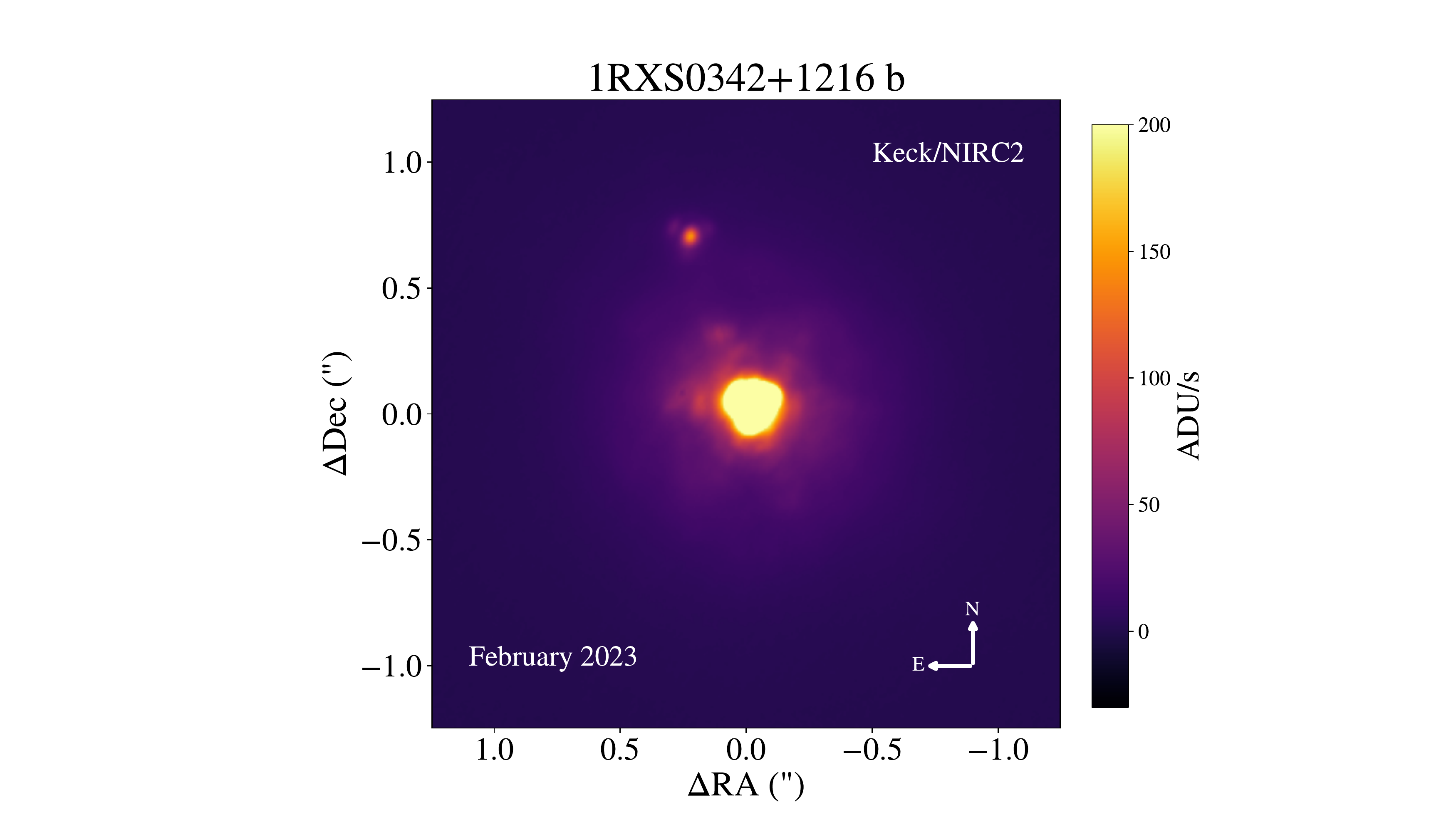}}
\caption{Final median combined image of the companion 1RXS0342+1216 b obtained with NIRC2 on Keck on 2023 February 8. }
\label{fig:nirc2median.png}
  \end{center}

\end{figure}

Our data reduction pipeline follows the prescription in \citealt{Yelda_2010}. We perform background subtraction, flatfielding, bad pixel correction, distortion correction using \texttt{Rain} (a Python adaptation of the distortion correction package \texttt{Drizzle} (\citealt{Fruchter_2002}) with the distortion solution from \citealt{Service_2016}, sub-pixel shifting to align the centroid of the primary in all frames, and finally a de-rotation of the image. For the de-rotation of the image, we use the parameters:
\begin{equation}
    \theta = \texttt{PARANG} + \texttt{ROTPOSN} - \texttt{INSTANGL} + \texttt{OFFSET}
\end{equation}
where the \texttt{PARANG} variable is the parallactic angle, \texttt{ROTPOSN} is the rotator position angle, \texttt{INSTANGL} is the instrument angle for NIRC2 and the \texttt{OFFSET} variable is the position angle offset needed to align the images with celestial north, measured to be 0.262 $\pm$ 0.018\textdegree  by \citealt{Service_2016}.
The final median combined image is shown in Figure \ref{fig:nirc2median.png}. In order to obtain astrometry of the companion and the primary, we use the StarFinder algorithm (\citealt{Diolaiti_2000}) for PSF centroiding. The astrometry is generated by centroiding the median of the images, while the uncertainties are determined by centroiding each individual image and then obtaining the standard deviation of the centroids. To convert from pixels to mas, we use the plate scale provided by \citealt{Service_2016} of 9.971 $\pm$ 0.004 mas/pixel. The final obtained relative astrometry of the companion is presented on Table \ref{tbl:astrometry}. Compared to previous astrometric data points from \citealt{Janson_2012}, \citealt{Janson_2014}, \citealt{Bowler_2015a},  \citealt{Bowler_2015b} and \citealt{Bowler_2020}, the companion appears to be moving towards the host star with an increasing position angle. 

\par

\begin{deluxetable*}{cccccc}
\tablecaption{Relative Astrometry of 1RXS0342+1216 b} \label{tbl:astrometry}
\tablewidth{20pt}
\tablecolumns{4}
\tabletypesize{\scriptsize}
\tablehead{\colhead{Epoch} & \colhead{Separation (mas)} & \colhead{Position Angle (\textdegree)} & Filter & Instrument & Reference}
\startdata
2007.951 &	883.0  ± 0.2 &	17.58 ± 0.09 & Ks & Keck/NIRC2 & \citealt{Bowler_2015b}\\
2008.63 &	860 	± 8	& 17.3 ±	0.4 &i' & AstraLux Norte & \citealt{Janson_2012}\\
2008.87 &	866 ± 8	& 17.8 ±	0.4 & i' &  AstraLux Norte & \citealt{Janson_2012}\\
2010.659 &	851	± 3	& 18.7 ±	0.1 & H &  Gemini-S/NICI & \citealt{Bowler_2015b}\\
2012.02	& 834 ±	57	& 17.6 ±	1.7 & i' & AstraLux Sur & \citealt{Janson_2014}\\
2012.645 & 831 ± 2	& 18.71 ±	0.07 & H & Keck/NIRC2 & \citealt{Bowler_2015a} \\
2013.044	& 822	± 8	& 19.1 ±	0.7 & L' & Keck/NIRC2 & \citealt{Bowler_2015a}\\
2018.08	& 772.3 ± 1.8 &	19.6 ±	0.1 & Ks & Keck/NIRC2  & \citealt{Bowler_2020}\\
2023.108 & 712.1 ± 0.4	& 20.82	± 0.15 & Ks & Keck/NIRC2 & This Work\\
\enddata
\end{deluxetable*}

\subsection{Keck/KPIC Data Reduction And Spectral Modeling} \label{kpic}
KPIC data for the companion was obtained on UT 2020 September 28 and 2021 July 3. For both epochs, the observations were done in K band (1.9 -- 2.4 $\mu$m) and have an average spectral resolution of R $\sim$ 35,000 (\citealt{Delorme2021}). The instrument presents a fiber injection unit with single-mode fibers which increase stellar right rejection and sky background \citep{Echeverri2022}. Our companion exposures for the 2020 epoch consisted of 6 exposures of 300 seconds each, with the companion in fiber 2, which presented the highest throughput after calibrations. We also obtained two host star exposures of 60 seconds each, with the host star present in fibers 1 and 2. For the 2021 epoch, we obtain 5 exposures of 300 seconds each for the companion, alternating between fiber 1 and fiber 2 to enable nod subtraction. The purpose of nodding, which is a strategy first implemented in 2021, is to reduce background noise, since the background taken during daytime calibrations may not match the night-time background. It is most useful for faint targets, so it does not significantly affect this target's SNR since it is fairly low contrast. We also obtain 2 exposures of the host star on fibers 1 and 2, with 60 second exposure for each, and exposures for an A0 star (HIP 16322) for telluric calibration. The data is reduced using the KPIC pipeline\footnote{\url{https://github.com/kpicteam/kpic_pipeline}} (\citealt{Wang_2021}). The reduction includes subtraction of adjacent frames for thermal background, bad pixel correction and wavelength calibration using data from the telluric star to fit the trace of the four fibers and nine orders. After this initial data reduction, we use the Python package \texttt{breads}\footnote{\url{https://github.com/jruffio/breads}} to analyze this dataset. \texttt{breads} follows the formalism of \citealt{Ruffio2019}, which uses a spline forward modeling methodology \citep{Ruffio2023}. The main challenge of high resolution spectroscopy is in the signal-to-noise ratio (SNR), where individual spectral lines present low SNR. In this method, the data is considered an addition of a model and a noise quantity:
\begin{equation}
  \bm{d}  = \bm{M}_{RV} \phi + \bm{n}
\end{equation}
    where the data vector \textbf{d} is composed of a model $\textbf{M}_{RV}$ combined with linear parameters $\phi$ and the noise vector \textbf{n}. The forward model relies on maximum likelihood estimation of the companion's signal by jointly fitting the companion data and the host star light from models for each of the components \citep{Agrawal_2022}. This way, the fit can account for starlight contamination at the companion's location. For the companion, the host star and the A0 telluric calibration star, we employ PHOENIX ACES AGSS COND models \citep{Husser_2013} as the forward model template. Unfortunately, the data from the 2021 epoch had poor throughput and therefore highly uncertain results in the atmospheric analysis of the companion and the host star due to poor observing conditions. For that reason, that epoch is not used for our analysis, and we only use KPIC data from 2020, which had a throughput of 1.4-1.8\%, depending on the fiber. 
\par
KPIC/NIRSPEC possesses nine spectral orders in $K$ band for each object. For this analysis, we only use orders 33 (2.29 -- 2.34 $\mu$m) and 32 (2.36 -- 2.41 $\mu$m), due to the high signal-to-noise and well-modeled telluric lines in these orders \citep{Ruffio2023}. Our algorithm follows an MCMC approach \citep{ForemanMacket2013}, and for all of our fits we use 512 walkers with 10,000 burn-in steps and 10,000 steps.
\section{Results} \label{results}

\subsection{Temperature, $log(g)$ and $[Fe/H]$} \label{templogg}

Using the \texttt{breads} package, we first use the 
Keck/KPIC data to obtain atmospheric parameters for both the companion and the host star. Our PHOENIX grids were used to fit for temperature, $log(g)$, metallicity ($[Fe/H]$) spin and radial velocity.

\begin{figure*}[ht]
    \centering
    \centering{{\includegraphics[width= 12cm]{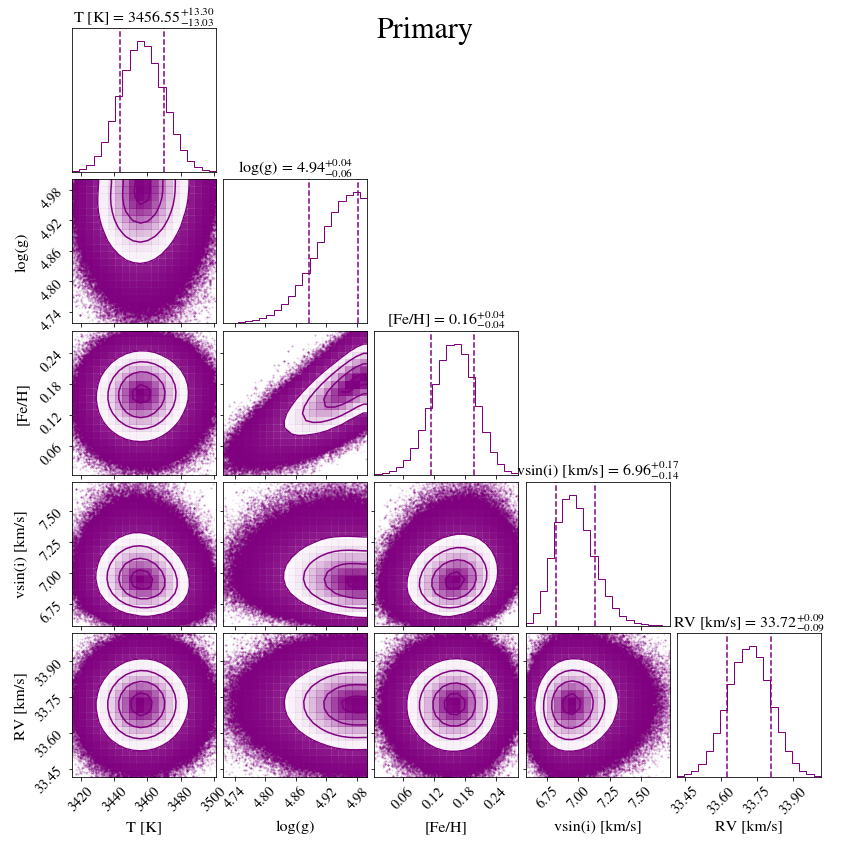} }}%
    \qquad
    \centering{{\includegraphics[width=12cm]{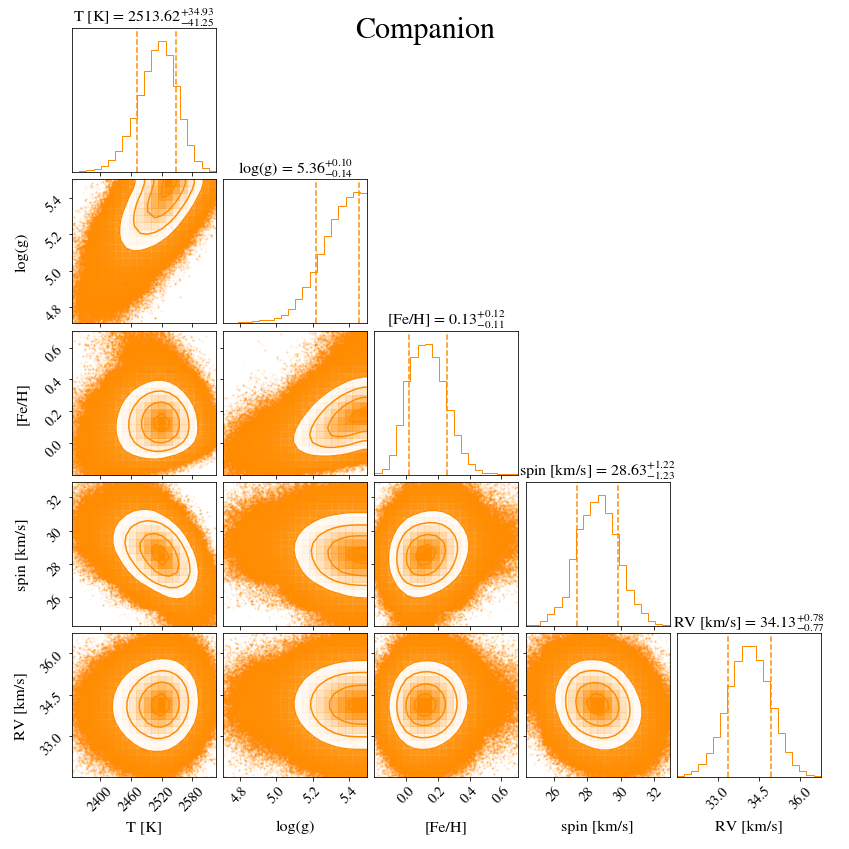} }}%
    \caption{Corner plots showing the posteriors for temperature, $log(g)$, $[Fe/H]$, $vsin(i)$ and RV for the primary and companion using KPIC data from 2020.}%
    \label{fig:kpiccorner}
\end{figure*}

\begin{figure*}[htb!]
  \begin{center}
\centerline{\includegraphics[width=19cm]{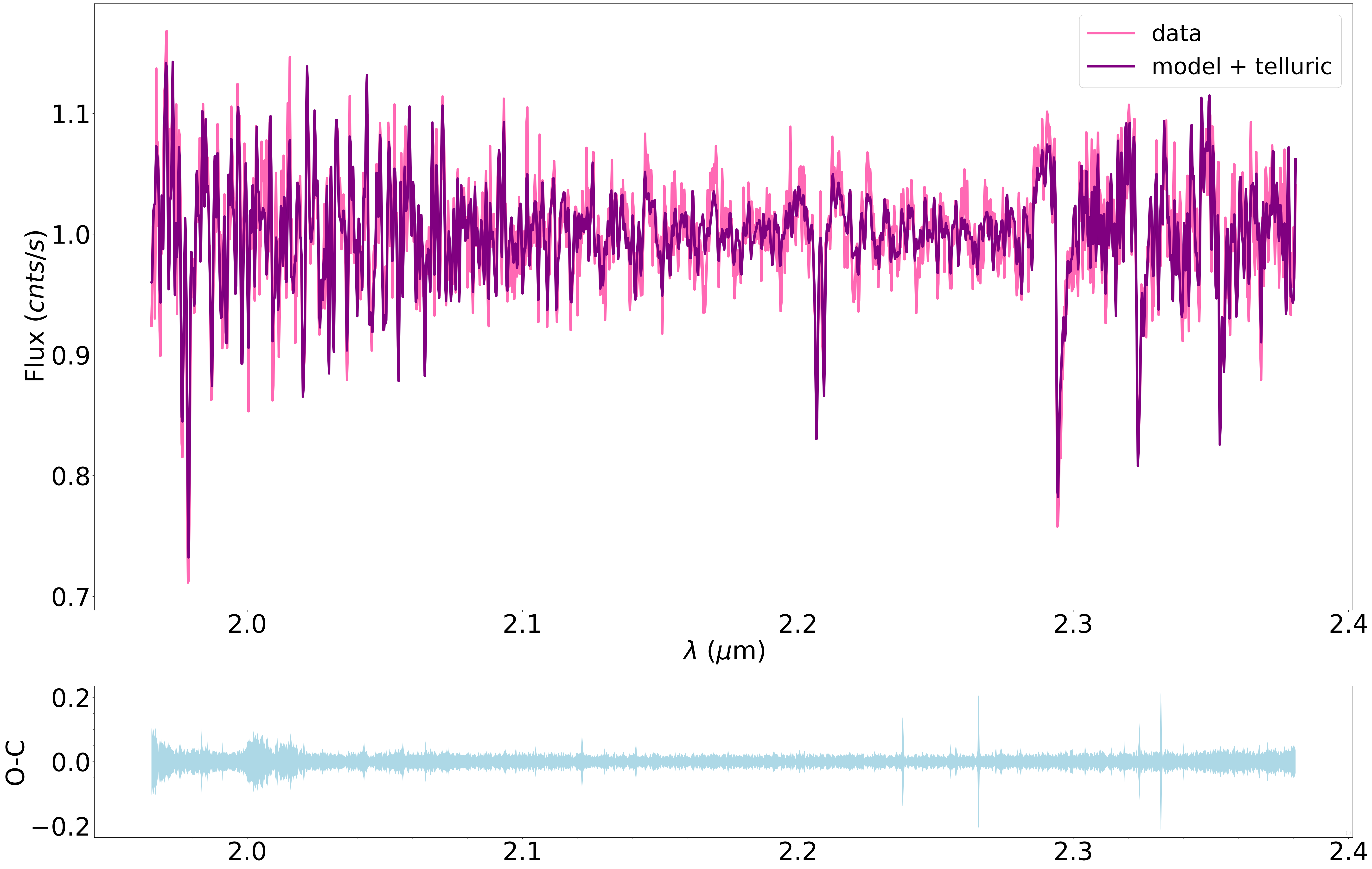}}
\caption{Final fit for the moderate resolution spectrum from OSIRIS. The temperature fit is 2760 $\pm$ 12 K, $log(g)$ is 5.2 $\pm$ 0.04. This pair of parameters yields a mass fit of 0.09 $\pm$ 0.02 $M_{\odot}$ for the companion, slightly above but still consistent with the KPIC fit. }
\label{fig:osiris}
  \end{center}

\end{figure*}

\begin{figure*}[htb!]
  \begin{center}
\centerline{\includegraphics[width=19cm]{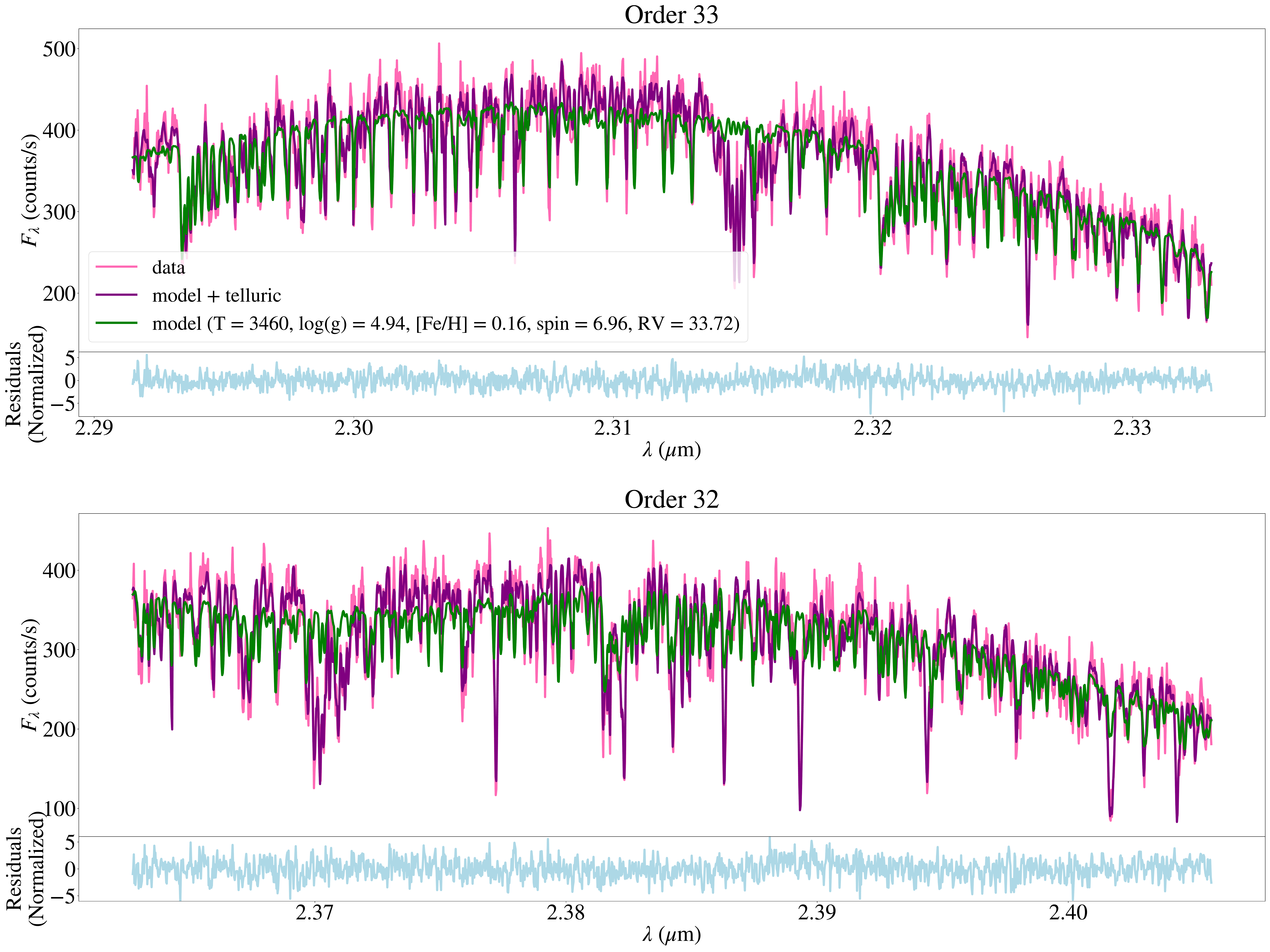}}
\caption{KPIC spectrum for the primary star compared to the best fit models from \texttt{breads}. The pink curve shows the data, with the green curve representing the atmospheric model and the purple curve representing the atmospheric model + telluric model. The blue curve in the lower panel shows the residuals, which are normalized by the data before uncertainty inflation.}
\label{fig:primary_spec}
  \end{center}

\end{figure*}

\begin{figure*}[htb!]
  \begin{center}
\centerline{\includegraphics[width=19cm]{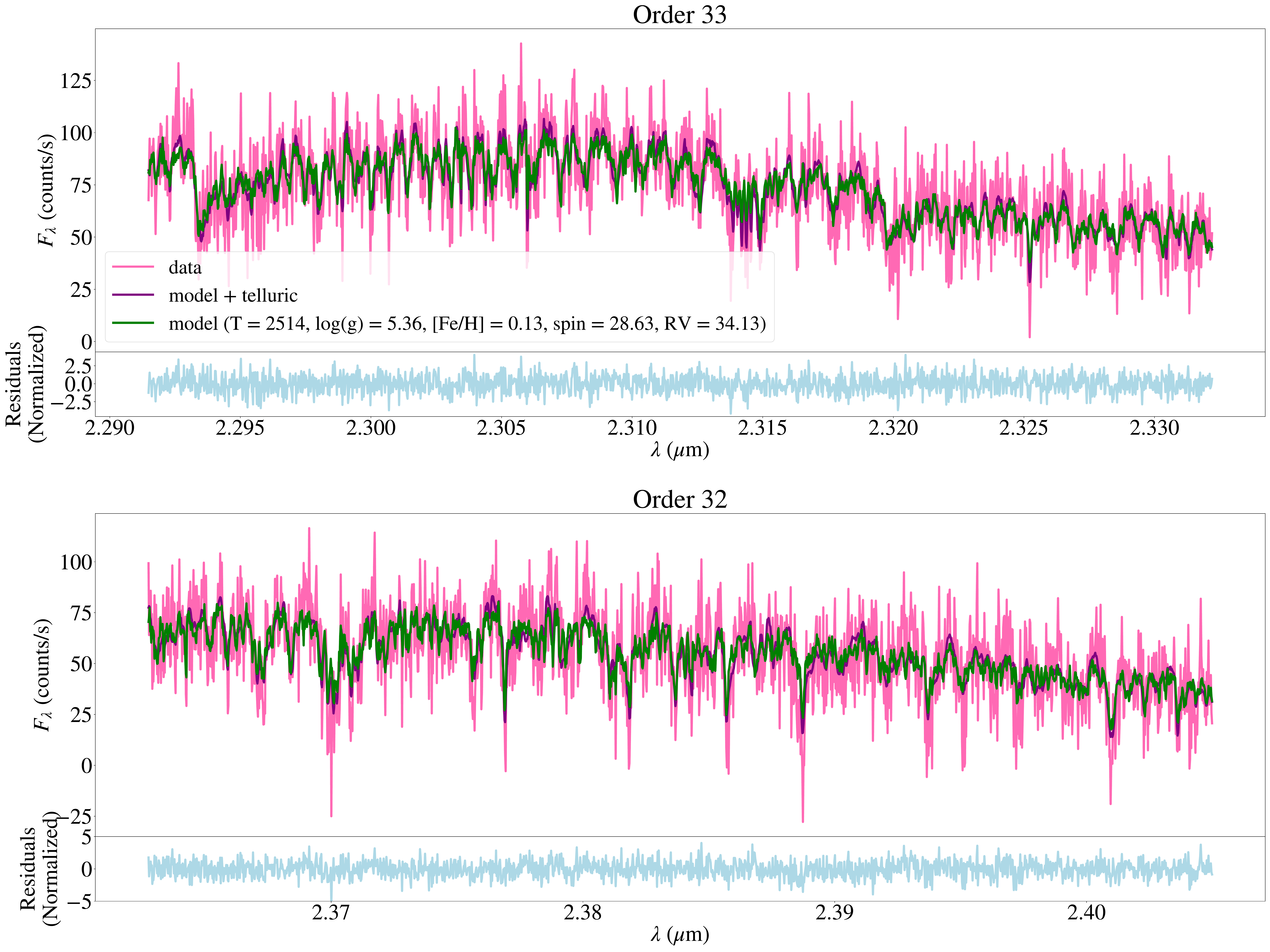}}
\caption{Same as Figure \ref{fig:primary_spec}, but for the companion.}
\label{fig:secondary_spec}
  \end{center}

\end{figure*}

Our final corner plots are presented in Figure \ref{fig:kpiccorner}.  We obtain a temperature of $2510 \substack{+40 \\ -40}$ K and a $log(g)$ of $5.36\substack{+0.10 \\ -0.14}$ (68th percentile) for the companion and a temperature of $3460 \substack{+10 \\ -10}$ K and a $log(g)$ of $4.94\substack{+0.04 \\ -0.06}$  for the host star. We note that the log(g) values for both objects are close to the grid ceiling, of 5.5 for the companion grid and 5.0 for the primary grid. The reason for these upper bounds is to keep the limit of log(g) in agreement with interior structure models, which we incorporate by considering the evolutionary models from \citealt{Baraffe_2015}. For any age in these models, values above $log(g) = $ 5.5 are inconsistent with the physical predictions. Surface gravity values that reach the grid ceiling have been reported by other works that use high resolution spectroscopy for isolated low-mass stars and brown dwarfs (e.g. \citealt{DelBurgo_2009}; \citealt{Hsu_2021}; \citealt{Hsu2024}), so this is a known model issue in atmosphere fitting to high resolution data. Other works have kept log(g) fixed in order to avoid this issue (e.g. \citealt{Blake_2010}; \citealt{Theissen_2022}).
\citealt{Bowler_2015a} obtained moderate resolution (R$\sim$4000) spectra of the companion using OSIRIS on Keck (0.365 -- 1.05 $\mu$m) .  In order to validate our atmospheric fits with KPIC data, we also fit the OSIRIS spectra using the same model grid. We fit for the temperature, $log(g)$ and RV using the forward-modeling routine Spectral Modeling Analysis and RV Tool (SMART; \citealp{Hsu_2021, Hsu_2021b}) using  PHOENIX COND models. We obtain a temperature of 2760 $\pm$ 12 K and a $log(g)$ of 5.2 $\pm$ 0.04 (68th percentile) with that dataset. The temperature of the companion is not consistent with the KPIC result (about 200 K higher), while the OSIRIS result yields a similar $log(g)$ value.  In this case, we fit for the data with the continuum subtracted from the spectrum. We present the OSIRIS best fit for the data in Figure \ref{fig:osiris}. We also plot our KPIC spectra compared to the models in Figures \ref{fig:primary_spec} and \ref{fig:secondary_spec}.  \par

The formal uncertainties from model fits for the atmospheric parameters are underestimated as they do not account for any systematic uncertainties in the models themselves. The formal statistical uncertainties from model fits for the atmospheric parameters are underestimated as they do not account for any systematic uncertainties in the models themselves or errors in interpolation of the grid between steps. In order to take this underestimation into account, we inflate our uncertainties  in temperature to $\pm$ 50 K and in $log(g)$ to $\pm$ 0.3 for both the companion and the host star, which corresponds to half of the model grid spacing size, of 100 K for effective temperature and 0.5 for $log(g)$. This uncertainty value is in accordance with our previous works that accounted for the underestimation in model uncertainties due to the systematics in grid interpolation (e.g., \citealp{Konopacky2013,Wilcomb_2020,Ruffio_2021,Hoch2022}). 

 \subsection{System Mass Derivation} \label{mass_section}
For this system,  \citealt{Bowler_2015b} reports a primary mass of 0.20 $\pm$ 0.05 $M_\odot$, a companion mass of
35 $\pm$ 8 $M_\mathrm{Jup}$ and therefore a total system mass of 0.23 $\pm$ 0.05 $M_\odot$. These masses were found by interpolating luminosity values with evolutionary models from \citealt{Baraffe1998} for the system's age, of 60 - 300 Myr. However, \citealt{Kuzuhara_2022} found using new Gaia data that this system is associated with the Hyades (membership probability $>$ 99.9\%), which present a much older age, at 750 $\pm$ 100 Myr \citep{BrandtHuang2015}. This older age estimate necessitates a recalculation of the mass of the primary and the companion in order to determine the total system mass. \par
In order to estimate the mass of the system given this new age estimate, we interpolate our newly derived effective temperature ($T_{\rm eff}$) and $log(g)$ chain pairs given by our KPIC 2020 data (see Figure \ref{fig:kpiccorner} for corner plots) with evolutionary models provided by \citealt{Baraffe_2015}, randomly sampling from the Hyades age estimate range in a Monte Carlo fashion to obtain a range of possible masses for the primary, secondary and the entire system.  \par
We obtain a median host star mass of 0.30 $\pm$ 0.15  $M_\odot$ and a median companion mass of 0.08 $\pm$ 0.01 $M_\odot$ (68th percentile). This adds to a total system mass of 0.38 $\pm$ 0.15 $M_\odot$. When we only include the effective temperature in our interpolation, we obtain masses that are consistent with the fit where $log(g)$ is included (0.41 $\pm$ 0.15  $M_\odot$ total system mass, with a primary mass of  0.33 $\pm$ 0.15  $M_\odot$ and secondary mass of 0.08 $\pm$ 0.01 $M_\odot$). Updating both the temperature and age of the components therefore yields higher masses. Our resulting mass for the companion places it near or slightly above the hydrogen-burning limit ($\sim$ 0.072 \Msol; \citealt{Burrows2001}), consistent with \citealt{Kuzuhara_2022}, who found 76 - 83 $M_\mathrm{Jup}$ using the companion's luminosity. We also fit for the mass of the companion using the temperature and $log(g)$ pairs from the OSIRIS best fit. We obtain a slightly higher but still consistent result of 0.09 $\pm$ 0.02 $M_\odot$ for the companion. \par

\subsection{Orbital Characterization with New Data} \label{masssys}

We use the new 2023 astrometry epoch to fit for the orbit of the companion. The most recent orbit fit of the companion was done by \citealt{DoO_2023} using observable-based priors, which have been shown to decrease biases in undersampled orbits done with direct imaging. The purpose of observable-based priors is to improve orbital estimates for orbits where the data covers a low  percentage of the orbital arc ($<$ 40\%). 
\par
We briefly summarize the formulation of observable-based priors. A detailed formulation is outlined in \citealt{ONeil_2019}. Observable-based priors assume that all regions of observable parameter space that can be observed are equally likely, emphasizing uniformity in observables rather than in model parameters (such as eccentricity and periastron passage epoch). Our orbit fit starts with measured observables from the astrometry, x(t), y(t), $v_z(t)$, 
where x and y are the object's positions ( right ascension (R.A.) and declination (Dec)) in the plane of the sky relative to the position of the primary ($x_o$ and $y_o$) and $v_z$ is the velocity relative to the star. These measured observables are linearly related to the orbital observables (which describe the position and motion in the orbital plane) by the Thiele-Innes constants (e.g
\citealt{Hartkopf1989}, \citealt{WrightHoward2009}). Due to this linear relationship, a uniform distribution in the measured observables would imply a uniform distribution in the orbital observables. The orbital observables, denoted here as X, Y, $V_x$ and $V_y$, are also connected to the model parameters according to the following equations (e.g. \citealt{hilditch_2001}; \citealt{Ghez2003}):
\begin{equation}
   X = a(cos E - e)
\end{equation}
\begin{equation}
   Y = a(\sqrt{1 - e^2}sin E)
\end{equation}
\begin{equation}
   V_x = - \frac{sin E}{1 - e cos E}\sqrt{\frac{GM}{a}}
\end{equation}
\begin{equation}
   V_y = \frac{\sqrt{1-e^2} cos E}{1 - e cos E}\sqrt{\frac{GM}{a}}
\end{equation}
where G is the gravitational constant, E is the eccentric anomaly, e is the eccentricity, a is the semimajor axis and M is the mass of the system. By transforming between measured observables and orbital observables, and then between orbital observables and model parameters, we can transform between measured observables and model parameters. This allows us to express a distribution that is uniform in the measured observables in terms of model parameters. Traditional uniform priors in orbital parameter space lead to a biased region in periastron passage ($T_o$), where the $T_o$ tends to artificially coincide with the observation epochs (\citealt{Konopacky2016}). This bias is mitigated in observable-based priors because they suppress this biased region of the parameter space when sampling it. \par

Here we revisit the analysis done in \citealt{DoO_2023} by fitting for the companion's orbital parameters with observable-based priors. We use the software Efit5 (\citealt{Mey2012}). Efit5 uses MULTINEST (\citealt{Fer2008}; \citealt{Fer2009}), a multimodal (or nested) sampling algorithm, to perform a Bayesian analysis on the data. For all of these fits, we use 3,000 live points in the nested sampling algorithm.

\begin{figure*}[ht!]
    \centering
    \centering{{\includegraphics[width= 8cm]{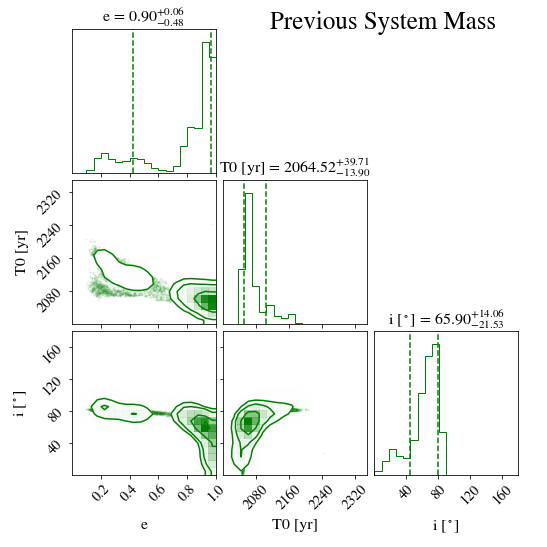} }}%
    \qquad
    \centering{{\includegraphics[width=8cm]{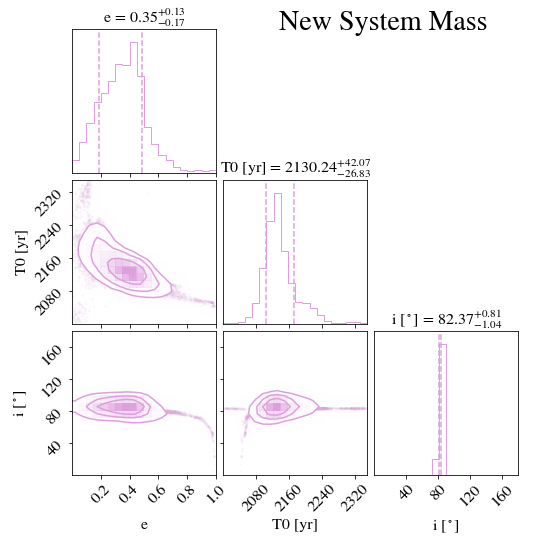} }}%
    \caption{Corner plots for 1RXS0342+1216 b's orbital parameters with the previous system mass of 0.23 $M_{\odot}$ (left) and new system mass of 0.38 $M_{\odot}$ (right). No new astrometry or radial velocity data were included in this orbit fit. Full corner plots can be found in Appendix \ref{cornerplots}.}%
    \label{fig:masscorner}
\end{figure*}

 Because orbit fits for directly imaged companions often rely on fixed total system masses to derive constraining posteriors when dynamical masses are not employed, we investigate how much the updated mass affects the original orbit fit for 1RXS0342+1216 b, by running the fit for both the old mass estimate (0.23 $M_\odot$) and new mass estimate (0.38 $M_\odot$). Since our interest in these first fits is only to see how the new mass affects the resulting orbital parameters, we include no new astrometry or RVs, and use only relative astrometry up to 2018 (see Table \ref{tbl:astrometry}). We use the distance estimate of 32.995 $\pm$ 0.0727 parsecs from Gaia DR2 (\citealt{GDR2}; as was done previously in \citealt{Bowler_2020} and \citealt{DoO_2023}).\par

\begin{figure}[htb!]
  \begin{center}
\centerline{\includegraphics[width=15cm]{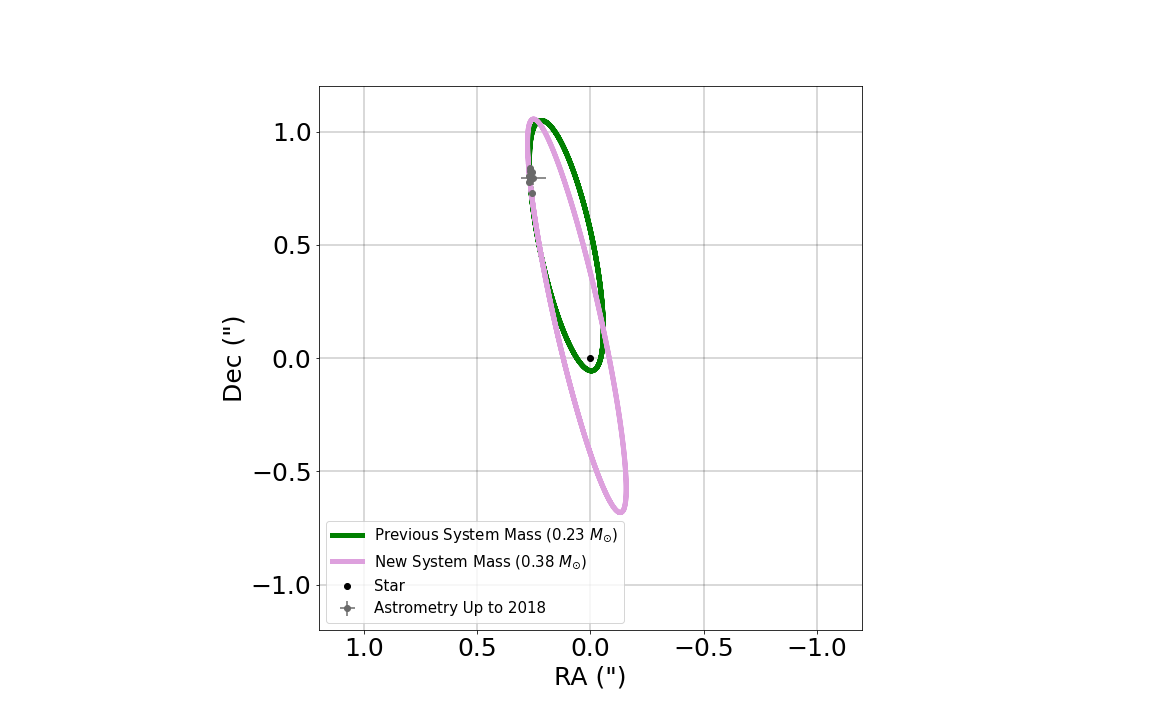}}
\caption{Visual orbit fit for the companion with the previous and new system mass. No additional astrometry/RV data were used in these fits.}
\label{fig:visualorbitmass.png}
  \end{center}

\end{figure}

 Our orbit fit results are presented in Figure \ref{fig:masscorner}. The orbit fit with the new mass estimate for the system significantly changes the orbital parameter posteriors. In particular, the eccentricity of the companion changes from $0.90\substack{+0.06 \\ -0.48}$ to $0.35\substack{+0.25 \\ -0.11}$, a much lower estimate. This change is coupled with a significant change in periastron passage, with the new mass estimate having wider spread of possible periastron passage epochs and placing it further away from current observation epochs. The inclination of the companion also changed from $65\substack{+10 \\ -40}$ $^{\circ}$ to $82\substack{+1 \\ -1}$ $^{\circ}$, moving from a highly uncertain orbit orientation to an orientation where the orbit is close to edge on. We show a visual example of these two orbit fits in Figure \ref{fig:visualorbitmass.png}. We also verify that a similar change in the eccentricity tail occurs with uniform parameter priors rather than with observable priors. Our results for uniform priors are presented in Appendix \ref{uniform}. \par

For all of our orbit fits with new data, we use an updated system distance of $32.96\substack{+0.022 \\ -0.024}$ pc from \citealt{BailerJones_2021}. We first fit for the orbit of 1RXS0342+1216 b using only the new astrometry epoch from 2023 obtained with Keck/NIRC2 and presented in Table \ref{tbl:astrometry}.   We find that the fit yields a period of $310\substack{+40\\ -40}$ years, with an eccentricity of $0.45\substack{+0.14 \\ -0.16}$. The last astrometric data point before the one presented in this work is from 5 years earlier, in 2018. We then fit for the orbit of the companion using only the relative radial velocity data from KPIC's 2020 dataset. The radial velocity of the host star, the companion and the relative radial velocity are presented in Table \ref{tbl:rvs}. From the measurement, the relative radial velocity of the companion is practically zero (0.41 $\pm$ 0.78 km/s). This signifies that the companion is not significantly moving towards or away from our line of sight at its current orbital phase, meaning that the degeneracy on the orbital plane remains mostly unconstrained (in particular the $\Omega$ and $\omega$ 180$^{\circ}$ degeneracy). Both of these resulting fits (new astrometry only and new RV only) can be found in Appendix \ref{cornerplots}. \par

\begin{deluxetable*}{ccccc}
\tablecaption{Relative Radial Velocity of 1RXS0342+1216 b} \label{tbl:rvs}
\tablewidth{20pt}
\tablecolumns{4}
\tabletypesize{\scriptsize}
\tablehead{\colhead{Epoch} & \colhead{Host Star RV} & \colhead{Companion RV} & \colhead{Relative RV (km/s)} & Instrument}
\startdata
2020.742 &	33.72 $\pm$ 0.09 & 34.13 $\pm$ 0.77  & 0.41 $\pm$ 0.78 & KPIC \\
\enddata
\end{deluxetable*}

 \begin{figure*}[ht!]
  \begin{center}
\centerline{\includegraphics[width=18cm]{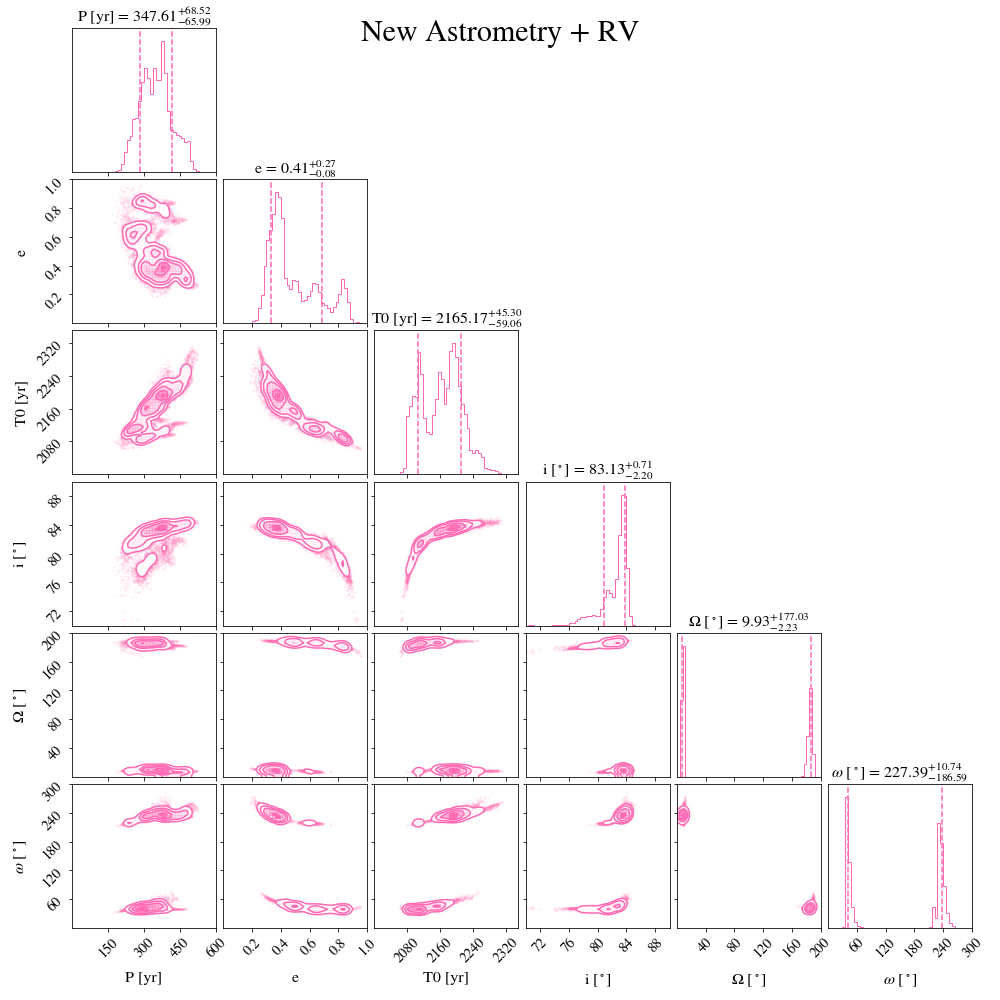}}
\caption{New orbit fit of 1RXS0342+1216 with the new 2023A astrometry epoch and 2020 RV data point.}
\label{fig:final_orbit_fit.png}
  \end{center}
\end{figure*}

\begin{figure*}[htb!]
  \begin{center}
\centerline{\includegraphics[width=18cm]{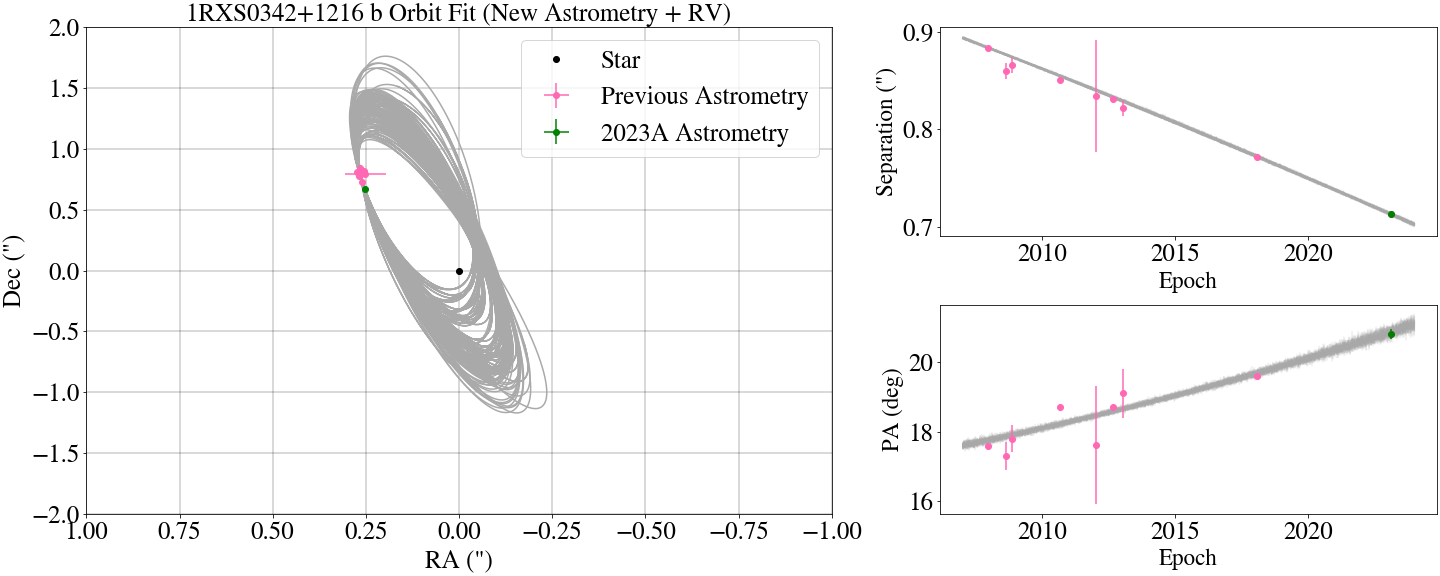}}
\caption{New orbit fit of 1RXS0342+1216 with the new 2023A astrometry epoch, the 2020 RV data point, and the updated total system mass.The left panel shows in gray 100 randomly sampled visual orbits from the posteriors with the current astrometry plotted on top (previous astrometry points are shown in pink while the 2023A astrometry point is shown in green). The top right plot shows in gray 100 randomly sampled orbits' separation in arcseconds from the host star as a function of epoch. The bottom right plot is the same as the top right, but with the position angle, or P.A., in degrees as a function of epoch.  }
\label{fig:orbitplot.png}
  \end{center}
\end{figure*}

Finally, we fit for the companion's orbit with the new total system mass, new astrometry epoch and with the RV obtained with KPIC. Here, we use both data types to maximize the amount of information given to the orbit fit. Our final orbit fit is presented in Figure \ref{fig:final_orbit_fit.png}.  We find that the companion's new orbit fit has a well constrained inclination of $83.13\substack{+0.71 \\ -2.20}$ $^{\circ}$, placing it in a nearly edge-on orbit. The eccentricity of the companion is $0.41\substack{+0.27 \\ -0.08}$, favoring moderate eccentricity solutions and difavoring circular ($<$ 0.2) and highly eccentric ($>$ 0.9) orbital solutions. The periastron passage of the companion is found to happen in $2165.17\substack{+45.30 \\ -59.06}$. This is about 150 years away from current observations. We plot 100 randomly sampled visual orbits and separation/position angle as a function of time in Figure \ref{fig:orbitplot.png}. \par
We also fit for the final orbit of the companion with the traditionally used uniform priors on the orbital parameters. The purpose of this exercise is to verify that both observable and uniform prior fits give similar orbital parameter posteriors. All of our uniform prior results are shown in Appendix \ref{uniform}. \par

The orbital posteriors for the different data and mass configurations are found in Table \ref{tbl:orbits}.

\begin{deluxetable*}{ccccccc}
\tablecaption{Orbit Fit for 1RXS0342+1216 b} \label{tbl:orbits}
\tablewidth{20pt}
\tablecolumns{7}
\tabletypesize{\scriptsize}
\tablehead{\colhead{Configuration} & \colhead{Period (years)} & \colhead{Eccentricity} & \colhead{T0 (year)} & \colhead{Inclination ($^{\circ}$)} & \colhead{$\Omega$ ($^{\circ}$)} & \colhead{$\omega$ ($^{\circ}$)} }
\startdata
Astrometry up to 2018, System Mass of 0.23 $M_{\odot}$  & $219.30\substack{+94.15 \\ -48.96}$ & $0.90\substack{+0.07 \\ -0.40}$ & $2057.69\substack{+29.11 \\ -9.58}$ & $64.99\substack{+14.10 \\ -39.28}$ & $180.98\substack{+13.69 \\ -157.25}$  & $91.33\substack{+157.40 \\ -73.51}$\\
Astrometry up to 2018, System Mass of 0.38 $M_{\odot}$ & $255.85\substack{+46.26 \\ -36.60}$ & $0.35\substack{+0.13 \\ -0.17}$ & $2130.24\substack{+42.07 \\ -26.83}$ & $82.37\substack{+0.81 \\ -1.04}$ & $13.04\substack{+178.19 \\ -2.88}$  & $213.00\substack{+19.12\\ -178.87}$ \\
Astrometry up to 2023, System Mass of 0.38 $M_{\odot}$ & $311.44\substack{+43.04\\ -44.70}$ & $0.45\substack{+0.14 \\ -0.16}$ & $2149.27\substack{+49.02 \\ -28.82}$ & $82.84\substack{+0.80 \\ -0.99}$ & $186.11\substack{+3.74 \\ -178.36}$  & $80.31\substack{+146.80\\ -35.89}$ \\
Astrometry up to 2018 + RV, System Mass of 0.38 $M_{\odot}$  & $253.72\substack{+72.21\\ -55.22}$ & $0.49\substack{+0.21 \\ -0.13}$ & $2118.59\substack{+49.22 \\ -34.03}$ & $81.78\substack{+1.25 \\ -2.49}$ & $9.31\substack{+118.40 \\ -1.73}$  & $218.50\substack{+11.41\\ -73.72}$ \\
Astrometry up to 2023 + RV, System Mass of 0.38 $M_{\odot}$ & $347.61\substack{+68.52\\ -65.99}$ & $0.41\substack{+0.27 \\ -0.08}$ & $2165.17\substack{+45.30 \\ -59.06}$ & $83.13\substack{+0.71 \\ -2.20}$ & $9.93\substack{+177.03 \\ -2.23}$  & $227.39\substack{+10.74\\ -186.59}$ \\
\enddata
\end{deluxetable*}

\subsection{Derivation of Atmospheric Parameters} \label{atmos}

\subsubsection{Fitting C/O Ratio} \label{CO}

In order to explore the C/O ratio of the companion, we generated a custom grid of PHOENIX models in which the abundances of carbon and oxygen were selected according to the predictions of \citet{Oberg_2011}, as described in \citet{Konopacky2013}.  Briefly, \citet{Oberg_2011} predict specific absolute abundances of C and O in the gas phase based on their model as a function of the ratio of solid to gas accretion in the atmosphere, which results in C/O between 0.45 and 1. The purpose of selecting specific values from this model was to generate a grid based on real predictions from physical models, and explore the likely values of C/O based on those models.  However, we interpolate between the models in the grid and therefore can explore any intermediate values of C/O.  We generated grids for both the host star and the companion in order to be able to compare the two. In our custom model grids, we hold the temperature, $log(g)$ and metallicity constant at the best fit values found in Section \ref{templogg} for the host star and the companion. Previous work has explored fitting the temperature and log(g) together with the C/O ratio in moderate resolution data (\citealt{Konopacky2013}). They find that fitting the temperature and log(g) simultaneously with C/O does not significantly affect the posteriors for C/O, but does require a significantly larger computational expense. For that reason, we do not fit for these parameters at the same time and instead fix temperature and log(g) when fitting for C/O. The grids are incorporated into the \texttt{breads} framework for fitting. 
\par
\begin{figure*}[ht]
    \centering
    \centering{{\includegraphics[width= 8.5cm]{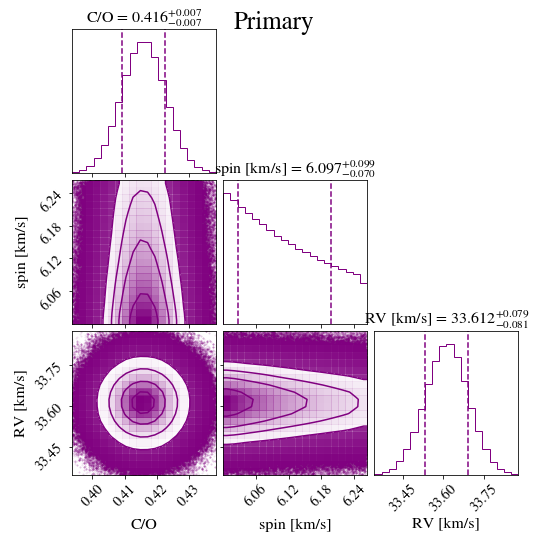} }}%
    \qquad
    \centering{{\includegraphics[width=8.5cm]{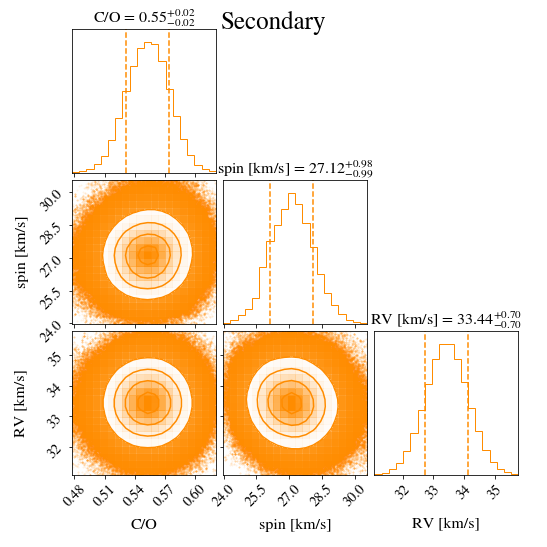} }}%
    \caption{Final C/O ratio fit for the primary and secondary. We also include a fit for the spin and RV using this model grid, and note that most values are broadly consistent with each other.}%
    \label{fig:COcorner}
\end{figure*}

The spacing in the C and O grid from \citealt{Oberg_2011} is not free but rather physically motivated - i.e it is derived from gas and grain abundances of C and O containing species (e.g. $CO_2$, CO, $H_2$O) and is parameterized over stellar abundances (from Figure 3 in \citealt{Oberg_2011}). Since we are performing a forward model rather than a free retrieval, these values must be chosen in advance. We chose the models presented by \citealt{Oberg_2011} because they are consistent with the expectations of C and O values for directly imaged objects and because the calculation is more computationally efficient when an already existing grid of C and O abundances is used.
When we fit for the C/O ratio for the primary, we find that the C/O values are closer to the lower boundary of the grid, indicating that its true C/O value is below 0.45. In order to remain consistent with their physical predictions, we incorporate lower C/O ratio values into the grid (down to the value of 0.25) by using a spline interpolation method to extend the C and O abundances derived in their work to lower values of C/O. With this extended grid, the resulting C/O ratios for the host star and companion are 0.416 $\pm$ 0.007 and 
0.55 $\pm$ 0.02, respectively (see Figure \ref{fig:COcorner}). The companion's C/O is broadly consistent with the solar value of $\approx$ 0.59 \citep{Asplund_2021}, while the primary has sub-solar C/O. The best fit C/O values yield small statistical error bars and slightly different, but still consistent spin/RV values than what was previously found using the PHOENIX model grid in Section \ref{templogg}. We note here that the C/O uncertainty value presented in the corner plots are not true uncertainties in the parameters, but rather statistical error bars due to the interpolation on the model grid. This interpolation generates systematic errors which do not reflect the true uncertainty of the C/O ratio for the object. Therefore, it appears that there are unaccounted uncertainties in the data that are not fully covered by the models. For this reason, we estimate that the C/O ratio found here is likely underestimating the measurement uncertainties, which are likely closer to 0.1 (e.g. \citealt{Hoch_2023}).

\subsubsection{Rotational Velocities} \label{rotations}

Our obtained $vsin(i)$ measurements are  $6.96\substack{+0.17 \\ -0.14}$ km/s and $28.63\substack{+1.22 \\ -1.23}$ km/s for the primary and secondary respectively. The upgraded NIRSPEC has R $\sim$ 35,000 (compared to the R $\sim$ 25,000 for the original NIRSPEC), which has an extrapolated minimum vsin(i) floor of 6.5 km/s, compared to 9 km/s for the original NIRSPEC (\citealt{Blake_2010}; \citealt{Hsu_2021}). The detection floor calculation for KPIC will be illustrated in an upcoming survey paper. The obtained vsin(i) measurement for the host star is close to the noise floor of the NIRSPEC instrument. For that reason, we conservatively adopt the noise floor of 9 km/s for NIRSPEC as an upper limit on the primary's spin for our assessment of rotation rates (i.e., the primary's $vsin(i) < $ 9 km/s.)\par
\citealt{Bowler_2023} has found using rotation period measurements from the primary that its orientation of the spin axis is most likely aligned with the plane of the orbit. In order to eliminate the degeneracy in inclination, we use spin axis inclination of the star obtained by \citealt{Bowler_2023} to calculate the rotational velocity of each object.
\par
We randomly sample from a Gaussian distribution with a mean of 66.5$^{\circ}$ and standard deviation of 10.8$^{\circ}$ for inclination measurements and from our $vsin(i)$ and inclination chains for the companion (we keep the upper limit of 9 km/s from NIRSPEC as the host's spin value) and obtain a distribution of rotational velocity of the companion and the host star. We find that the host star and the companion have very different rotational speeds, with $V_{host}$  $\lesssim$ 9 km/s and $V_{companion}$ = $29.0\substack{+3.5 \\ -3.5}$ km/s, yielding a ratio of $V_{companion}$/$V_{host}$ of $\gtrsim$ 3. The companion is therefore spinning at least 3$\times$ faster than the host star. \par
Break-up velocities can be calculated for an object with a given radius and mass using the equation:
\begin{equation}
    v_{break} = \sqrt{\frac{GM}{R}}
\end{equation}
where G is the gravitational constant, M is the object's mass and R is the object's radius (e.g. \citealt{Konopacky2012}, \citealt{Wang_2021}). We use this equation to estimate the break-up velocity of the primary and secondary. In order to perform this calculation, we estimate the radius of the targets using the same procedures to estimate the mass from evolutionary models by \citealt{Baraffe_2015} (i.e., interpolating over the chains for temperature and $\log(g)$). We obtain radii of 0.39 $\pm$ 0.10 $R_\odot$ for the primary and 0.11 $\pm$ 0.08 $R_\odot$ for the secondary. Coupled with the masses obtained in section \ref{masssys}, we obtain break-up velocities of 383 $\pm$ 64 km/s and 373 $\pm$ 24 km/s respectively. The two objects present similar break-up velocities, but the companion spins significantly faster than the host, at slightly above 8\% of its break-up velocity, consistent with \citealt{Bryan_2020}, which expects young companions to have spins at less than 10\% of their break-up speed. \par

\subsection{Summary of System Properties}

\begin{deluxetable*}{ccc} [ht!]
\tablecaption{Final System Properties} \label{tbl:properties}
\tablewidth{20pt}
\tablecolumns{3}
\tabletypesize{\scriptsize}
\tablehead{\colhead{Property} & \colhead{Host Star (Primary)} & \colhead{Companion (Secondary)}}

\startdata
Temperature (K) & 3460 $\pm$ 50 & 2510 $\pm$ 50 \\ 
$log(g)$ & 4.94 $\pm$ 0.30 & 5.36 $\pm$ 0.30  \\
$[Fe/H]$ & 0.16 $\pm$ 0.04 & $0.13\substack{+0.12 \\ -0.11}$ \\
v sin(i) (km/s) & $<$ 9 km/s & $28.63 \substack{+1.22 \\ -1.13}$ \\
Mass & 0.30 $\pm$ 0.15 & 0.08 $\pm$ 0.01 \\
Radius ($R_\odot$) & 0.39 ± 0.10 & 0.11 ± 0.08 \\
Break Velocity (km/s) & 383 ± 64 & 373 ± 26 \\
C/O & 0.42 $\pm$ 0.10 & 0.55 $\pm$ 0.10 \\
\enddata
\tablecomments{The temperature, $log(g)$ and C/O ratio present the inflated uncertainties. The mass, radius and consequently break velocity are derived by interpolating temperature and $log(g)$ measurements with evolutionary models from \citealt{Baraffe_2015}.}
\end{deluxetable*}

We summarize the final derived system properties using Keck/KPIC data and Keck/NIRC2 data for the 1RXS J034231.8+121622 system. The properties are presented in Table \ref{tbl:properties}.

\section{Discussion} \label{disc}

\subsection{Temperature, $log(g)$ and System Mass}
Our resulting fits for KPIC data provide us with an insight on the host star and companion's temperature, $log(g)$ and $[Fe/H]$. The final properties of these objects allow for a new system mass fit using evolutionary models. One particularly important change from previous results in the mass estimate for this system is the likely placement of this system as part of the Hyades found by \citealt{Kuzuhara_2022}. The older age of 750 $\pm$ 100 Myr of the Hyades (\citealt{BrandtHuang2015}) places both the host and the companion at higher mass estimates than previous measurements, which assumed an age of 60 -- 300 Myr. The host star's temperature and $log(g)$ makes it about 60\% more massive than the previous estimate, while the companion is about 39\% more massive than with previous measurements. The mass ratio (companion/host), which was $\approx$ 0.17, is now closer to 0.28. \par
For the companion, this new mass estimate places it near or above the hydrogen-burning limit, with the possibility of it being a low-mass star rather than a brown dwarf. Therefore, questions about formation of this system's secondary can be answered using a variety of parameters, such as its orbital parameters (in particular the eccentricity) and atmospheric parameters. \par

\subsection{Orbit Fit}

The orbit fit of the companion presented significant changes once the new estimate in system mass, the NIRC2 astrometry and KPIC RV data points were incorporated. Without the addition of new data, the orbital fit had a large tail of high eccentricity and low inclination orbits approaching periastron passage. These posteriors are a result of a known bias when the orbital information is undersampled and the fit is prior dominated (e.g. \citealt{ONeil_2019}, \citealt{DoO_2023}, \citealt{Blunt2023}). In particular, the eccentricity-inclination bias shown in \citealt{FerrerChavez2021} appears to apply to this system's orbital posteriors, where it becomes difficult to distinguish between face-on, eccentric orbits and edge-on, circular orbits. \par
The change in system mass, however, also played a significant role on the final orbital posteriors. The possibility of major changes in orbital parameters with changing mass is briefly addressed by \citealt{Bowler_2020}, who found that some companions in their orbit fits (including the one in this work) presented posteriors with a peak of eccentricities near 1.0 that could be caused by the wrong system mass estimate or underestimated astrometric uncertainties. Indeed, in this case no new astrometric or radial velocity data were needed to significantly change the eccentricity posterior of the companion (and other posteriors as well, such as inclination and $T_0$). The reason for such a large difference in the companion's orbit when changing the mass is Kepler's Third Law: for a more massive system, the orbital velocities of the objects are also increased (i.e., they orbit faster around each other).
 Previously, the fast changing astrometry likely yielded an orbital velocity that, with a smaller system mass, could only be explained by an object that is at the fastest point in its orbit, i.e. the periastron passage. This required the orbit to be more eccentric and thus more ``face on" relative to our line of sight. \par
 This result implies that the undersampling and uncertainty in ``visual" orbital information, such as astrometry, is not the only contributor to the limitations in orbital characterization of directly imaged companions. Uncertainties in system age, distance, and consequently mass, which are all assumed to be known and held fixed or fit with a Gaussian prior in orbit fitting packages (\citealt{ONeil_2019}, \citealt{Blunt_2020}, \citealt{Brandt_2021}), also play a major role in the orbital determination of these objects and may be highly uncertain. \par
\par

The new KPIC RV data point is near zero, and therefore it is not sufficient to eliminate the degeneracy in $\Omega$ and $\omega$, which would more accurately determine the orbital plane of the companion. However, it still provides valuable information for the orbit fit. Together with the new system mass, the new data places the orbit at a well-defined inclination of $\sim$ 83 $^{\circ}$, with the periastron passage occurring about 142 years from the last observation. The period is $\approx$ 350 years. In Appendix \ref{uniform}, we present our uniform prior results, which are consistent with the observable prior results. \citealt{DoO_2023} found that the minimum orbital coverage required to obtain reliable orbital parameter posteriors is at about 15\% of the orbital period. Therefore, despite the consistency in results regardless of prior choice, the orbit is still undersampled at about 4.3\% of orbital period coverage. For that reason, there is a possibility that the orbital posteriors presented here will change with the addition of new data. It is also possible that the inclusion of radial velocity data can decrease the coverage value needed to leave the prior-dominated regime, however, that possibility is yet to be tested and should be explored in future work.

\subsection{Metallicity}

The metallicities for the primary and for the companion are 0.16 $\pm$ 0.04 and $0.13\substack{+0.12 \\ -0.11}$ respectively. These metallicities indicate abundances slightly above the Solar value of 0.012 \citep{Asplund2006}. These values also hint that this system likely formed in a region that is more metal rich. Indeed, the metal enrichment of the Hyades ($[Fe/H]$ varies between 0.11 -- 0.21; e.g. \citealt{Perryman1998}, \citealt{Maderak2013}, \citealt{Brandner2023}) further contributes to the likely association of this system with the cluster stipulated by \citealt{Bowler_2015b} and \citealt{Kuzuhara_2022}. We repeat the procedure performed by \citealt{Kuzuhara_2022} using the BANYAN-$\Sigma$ algorithm (\citealt{Gagne2018}) with the updated radial velocity for the system from KPIC and obtain a Hyades membership probability of 99.8\%. The new found Hyades membership also imposes for better constraints on age and distance to the system, which improve both mass and orbit estimates for the objects. \par

\subsection{C/O Ratio}

The C/O ratios obtained for the host star and the secondary are 0.42 and 0.55, respectively. Utilizing uncertainties that account for potential systematics in the models themselves of $\pm$ 0.1, these two values are broadly consistent with each other (about $1\sigma$ away), and the primary's value is broadly consistent with the Sun's C/O ratio of 0.59 found by \citep{Asplund_2021}.  Despite the fact that the new larger mass of the companion means that this source could not have formed via core accretion (see Section \ref{formationpath}), there are two reasons why this measurement is still helpful.  First, it provides a means for verifying the C/O ratios of other sources (e.g., \citealt{Phillips2023}).  If the measured values had been significantly different from each other or from the solar value, it may have called into question measurements for planets in which C/O is not near 0.59.  Thus systems such as this one verify the expectation of chemical homogeinity between companion and star and provide important cross-checks for objects with similar spectral features that may have formed via core/pebble accretion.  Secondly, it places this object in the broader context of other systems in which C/O is being measured, allowing for the investigation of potential values of C/O across the full mass spectrum of measured companions. Additional data points at higher masses will help to more strongly identify any breaks in C/O as a function of properties such as mass, separation and/or age (e.g., \citealt{Hoch_2023}).

%The C/O ratios obtained for the host star and the %secondary are 0.47 and 0.57 respectively. Despite the %underestimated uncertainties, both values do not %indicate an enhanced C/O ratio. In fact, the %similarity of the C/O ratio of these objects with %solar values and with each other's values indicates %that the companion is unlikely to have formed in a %%core-accretion scenario. Distinguishing between %gravitational instability and binary star formation %mechanisms is a difficult task using C/O alone. 
%qmk write here

\subsection{Rotational Velocity}
The objects' rotational velocities were assessed assuming alignment of the rotational axis with the orbital plane. We find that the objects have very different spin velocities, of at most 9 km/s for the host star and $\sim$ 29 km/s for the companion. %The host star's rotational velocity is at the upper limit found by \citealt{Bowler_2023} of 7 km/s.  
The companion spins at least $\sim$ 3x faster than the host star, which is consistent with the theoretical predictions of magnetic braking timescales for lower mass objects (e.g. \citealt{West2008}). The longer timescales for magnetic braking laws cause objects of lower mass to often be rapid rotators (e.g. \citealt{Konopacky2012}), and can explain the discrepancy in rotational velocities of the host and the companion. For instance, \citealt{Reiners2008} found that mid M-dwarfs are slow rotators while late M-dwarfs are fast rotators, with the rotation speed being strongly correlated with an object's magnetic field. \citealt{West2008} quantified this process by generating an age-activity relationship for M dwarfs. The temperature of the primary places it in an early M spectral type. The companion's temperature places it in a late M spectral type \citep{Pecaut_2013}. With these spectral types, \citealt{West2008} predicts an activity lifetime of about 2 $\pm$ 0.5 Gyr for the primary and over 8 Gyr for the secondary.   After estimating the break velocities of both objects, the companion is only spinning at $\sim$ 8\% of its break velocity, which is less than a few objects found by \citealt{Konopacky2012} for instance, which were spinning at $\sim$ 30\% of their break-up speed.  \par

\par

\subsection{Formation Pathways} \label{formationpath}

%This system is comprised of a M dwarf host star with a massive companion right at the hydrogen burning limit. The C/O ratios of both components are similar to solar values, disfavoring the formation of the companion object via core accretion. This is in agreement with other works, such as \citealt{Schlecker2022}, who found using simulations that stars below 0.5 \Msol cannot produce a population of giant planets.  Several other theoretical scenarios have also found it harder to build objects larger than $\sim$30 M$_{Jup}$ via core or pebble accretion. 

The RXS J034231.8+121622 system is comprised of an M dwarf host star with a massive companion at or about the hydrogen-burning limit. The C/O ratios of both components are similar to each other, with no C/O enhancement in the companion. This is expected since theoretical scenarios predict difficulty in forming objects above $\sim$30 M$_{J}$ via core or pebble accretion, particularly around low mass stars (e.g., \citealt{Mordasini2012,Schlecker2022}).   However, sources such as 1RXS0342+1216 b may be candidates for formation via gravitational instability in a disk.  Using the predictions of \citet{Kratter2010} for the potential mass of an object that can be formed via disk instability, we find that companions as large as 153 M$_{Jup}$ are feasible at the best-fit semi-major axis of $\sim$36 AU.  Thus, exploring whether higher mass companions formed via gravitational instability in a disk versus a binary star formation scenario is an interesting area of investigation, especially in the context of understanding the full spectrum of outcomes for the companion formation process. 
\par
However, distinguishing between disk instability and binary star formation mechanisms (e.g. disk fragmentation) is a challenging task.  For example, the eccentricity of the companion is moderate, at about 0.4. Despite the potential usefulness of the eccentricity parameter as a formation tracer at a population level, it is difficult to determine formation pathways for an individual system by assessing the eccentricity alone, because gravitational instability has been found to produce clumps of eccentricities as high as 0.35 in simulations (\citealt{Mayer_2004}), and binary star formation studies have found that close binary stars of $<$ 100 AU separation have uniform population eccentricity distributions \citep{Hwang2022}.  \par

\section{Conclusion} \label{conc}
This work characterizes the 1RXS J034231.8+121622 system using high resolution spectroscopy from the Keck Planet Imager and Characterizer (KPIC) and a new astrometry data point from Keck/NIRC2. The main findings of this study are:

\begin{itemize}
    \item We find the temperature, $log(g)$, $[Fe/H]$, spin, RV and C/O ratios for both the host star and the companion. The relative RVs, spins and C/O ratios are reported for the first time for this system (see Table \ref{tbl:properties}).
    \item We use the temperature and $log(g)$ posteriors to re-derive the masses of these objects using an updated age for the system, which is now older than previously derived due to its likely association with the Hyades (750 $\pm$ 100 Myr). The $[Fe/H]$ measurement is well in agreement with the Hyades value. The masses of both objects increased, by 60\% for the host star and by 39\% for the companion.
    \item We find that the system mass, which is generally taken as a fixed parameter or fit for using a Gaussian prior in orbit fits, changes our orbital parameter posteriors substantially even without the addition of new astrometry or radial velocities. The increase in system mass causes the periastron passage of the companion to be further away in time from observations.
    \item The orbital parameter posteriors for the companion hint at a moderate eccentricity of $\sim$ 0.4 - 0.5, a result which appears to be independent of priors. The tails of low eccentricities ($<$ 0.2) and high eccentricities ($>$ 0.9) are now disfavored by the addition of new data. However, the eccentricity distribution is still uncertain and will likely require further observations to better constrain it.
    \item The C/O ratios for the host and the companion are 0.42 $\pm$ 0.10 and 0.55 $\pm$ 0.10.  Both values are broadly consistent with solar values. 
    \item Previous works have found that the companion and the host spin axis are likely aligned with the orbital axis. If this is the case, the companion is spinning at least 3x faster than the host, as is expected for lower mass objects. The companion spins at about 8\% of its break-up velocity.
    \item From the eccentricity, mass ratio and C/O ratio of the objects, the formation of this system did not occur from a core accretion scenario. Whether it occurred via gravitational instabilities in a protoplanetary disk or disk fragmentation in a protostellar disk remains unclear.

\end{itemize}

This work shows that high resolution spectroscopy is a powerful tool for characterizing directly imaged systems. The high resolution spectra allowed for precise radial velocity measurements of substellar companions and for atmospheric parameter estimation such as temperature, $log(g)$, $[Fe/H]$, spin and C/O ratio. Together, the detailed characterization of these systems provide clues on the formation processes of these companions, both individually and at a population level.
\section{Acknowledgements}
We thankfully acknowledge Brendan Bowler for sharing OSIRIS data on the 1RXS J034231.8+121622. We also thank the anonymous referee for providing comments that helped improve this manuscript.\par

Some of the data presented herein were obtained at the W. M. Keck Observatory, which is operated as a scientific partnership among the California Institute of Technology, the University of California, and the National Aeronautics and Space Administration. The W. M. Keck Observatory was made possible by the financial support of the W. M. Keck Foundation. The authors wish to acknowledge the significant cultural role that the summit of Maunakea has always had within the indigenous Hawaiian community. The author(s) are extremely fortunate to conduct observations from this mountain. Portions of this work were conducted at the University of California, San Diego, which was built on the unceded territory of the Kumeyaay Nation, whose people continue to maintain their political sovereignty and cultural traditions as vital members of the San Diego community. \par
C.D.O. is supported by the National Science Foundation Graduate Research Fellowship under Grant No. DGE-2038238. Further support for this work at UCLA was provided by the W. M. Keck Foundation, and NSF Grant AST-1909554. J.X. acknowledges support from the NASA Future Investigators in NASA Earth and Space Science and Technology (FINESST) award \#80NSSC23K1434. Any opinions, findings, and conclusions
or recommendations expressed in this material are those of the author(s) and do not necessarily reflect
the views of the National Science Foundation.

Funding for KPIC has been provided by the California Institute of Technology, the Jet Propulsion Laboratory, the Heising-Simons Foundation (grants \#2015-129, \#2017-318, \#2019-1312, and \#2023-4598), the Simons Foundation (through the Caltech Center for Comparative Planetary Evolution), and the NSF under grant AST-1611623.

Part of this work was carried out at the Jet Propulsion Laboratory, California Institute of Technology, under contract with NASA (80NM00018D0004).

\clearpage
\bibliographystyle{aa_bst.bst}
\bibliography{main.bib}
\clearpage
\appendix

 \section{Full Corner Plots} \label{cornerplots}

In this section we present the full corner plots for the companion's orbit fit with different amounts of data.
\begin{figure}[ht!]
    \centering
    \centering{{\includegraphics[width= \columnwidth]{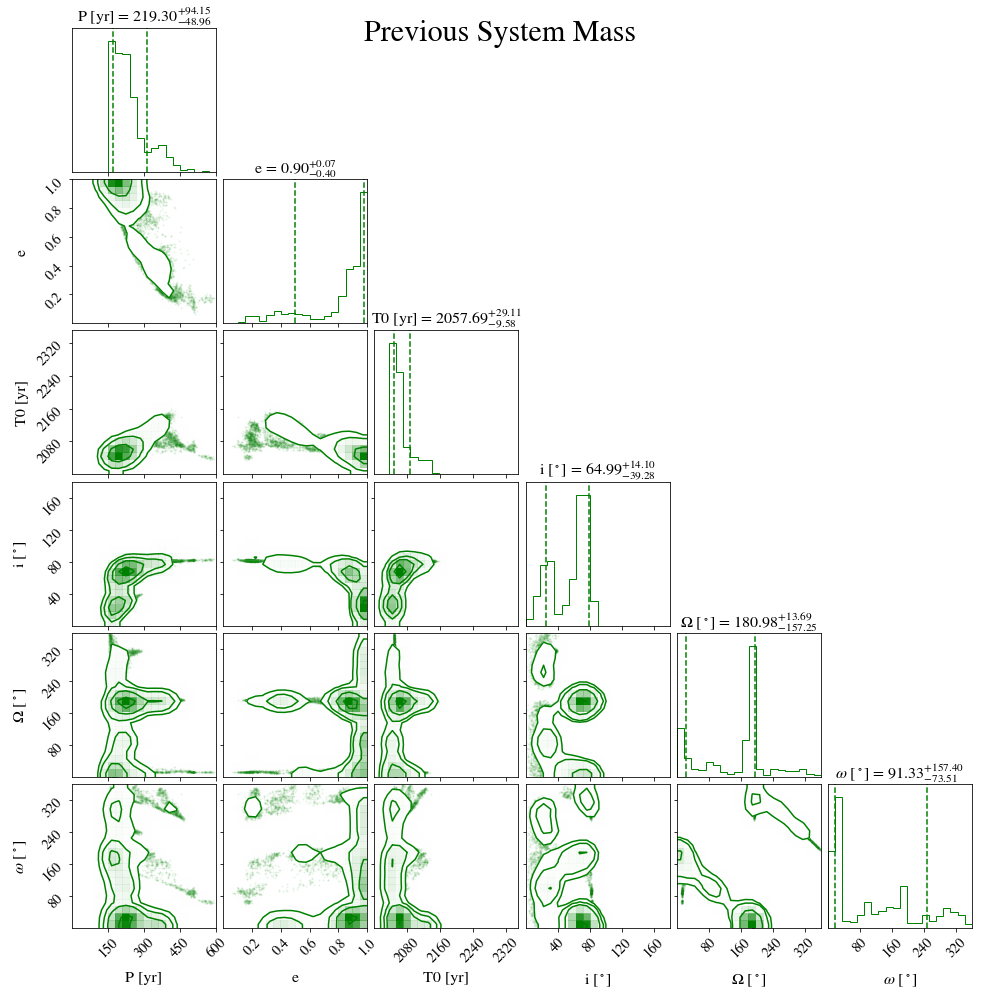} }}%
    \caption{Corner plots for 1RXS0342+1216 b's orbital parameters with the previous system mass of 0.23 $M_{\odot}$. No new astrometry or radial velocity data were included in this orbit fit.}%
    \label{fig:newmasscorner}
\end{figure}

\begin{figure}[ht!]
    \centering
    \centering{{\includegraphics[width= \columnwidth]{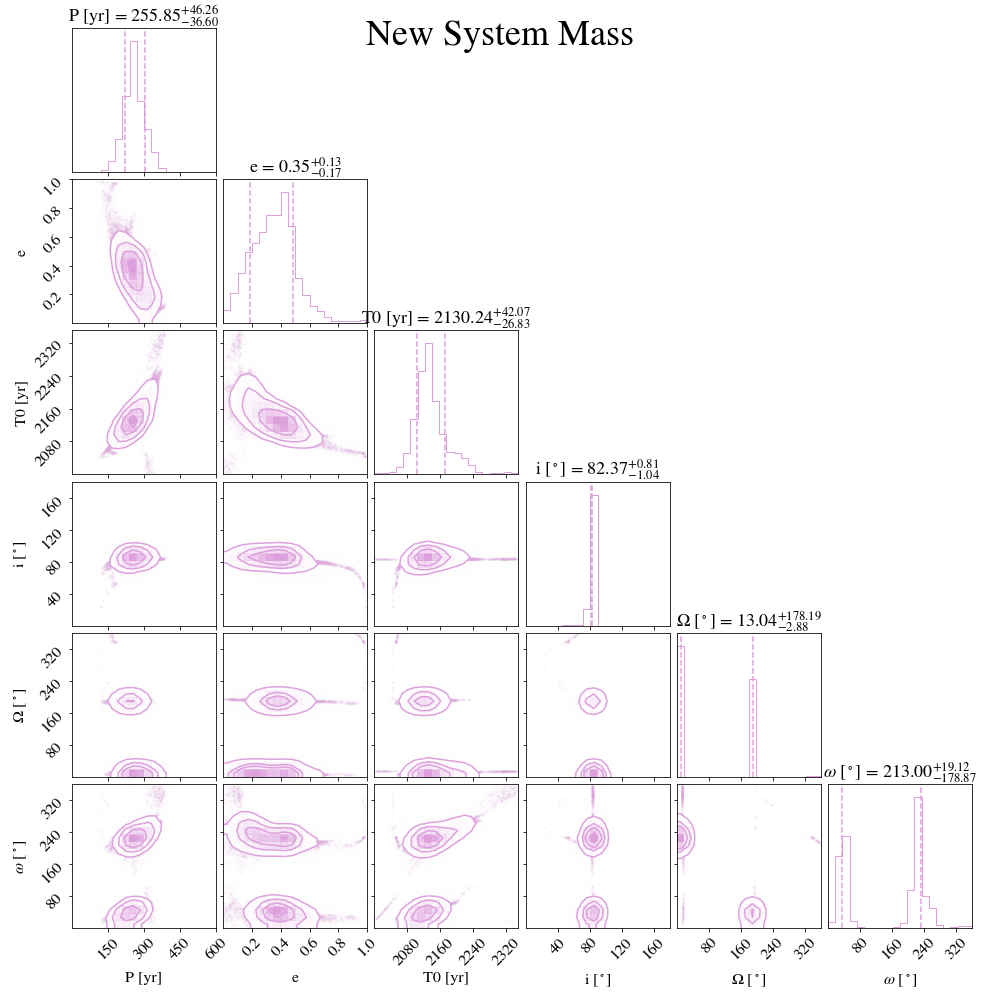} }}%
    \caption{Corner plots for 1RXS0342+1216 b's orbital parameters with the new system mass of 0.38 $M_{\odot}$. No new astrometry or radial velocity data were included in this orbit fit.}%
    \label{fig:oldmasscorner}
\end{figure}

 \begin{figure}[htb!]
  \begin{center}
\centerline{\includegraphics[width=\columnwidth]{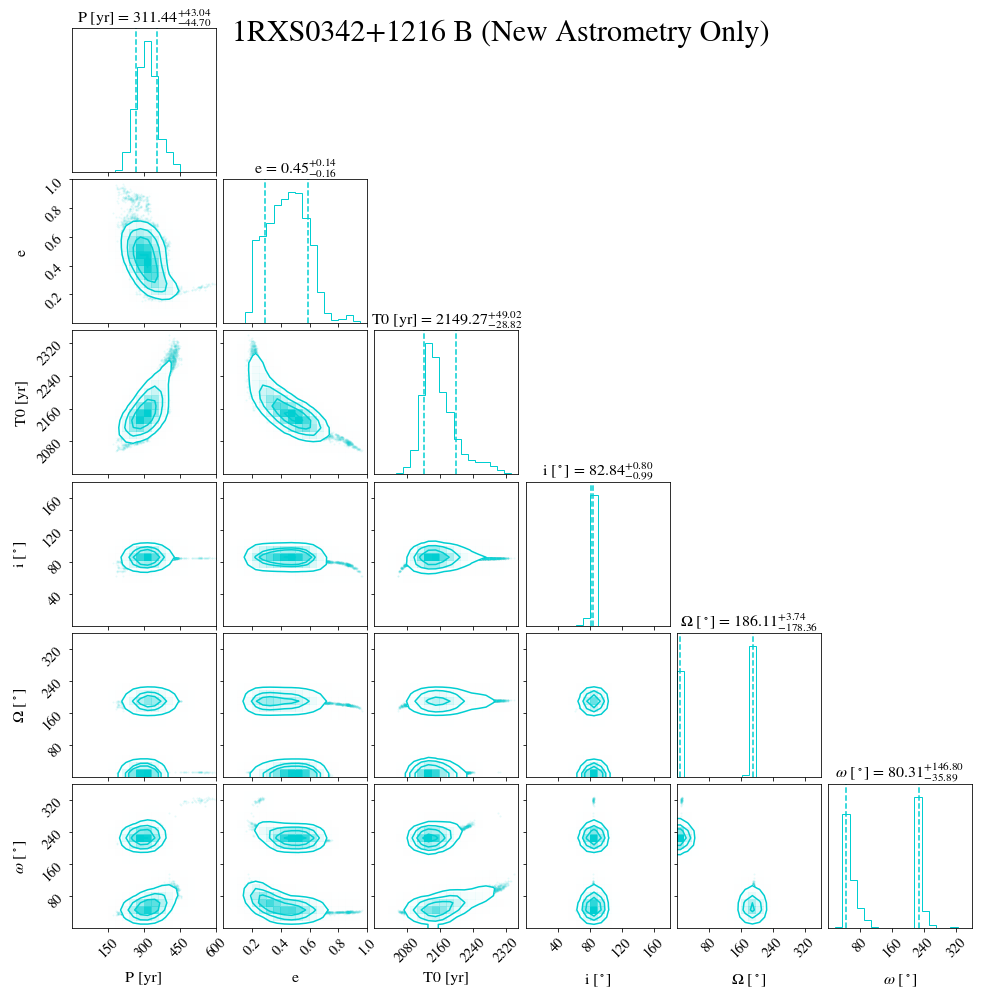}}
\caption{New orbit fit of 1RXS0342+1216 with the new 2023A astrometry epoch.}
\label{fig:orbit_astroonly.png}
  \end{center}
\end{figure}

\begin{figure}[ht!]
  \begin{center}
\centerline{\includegraphics[width=\columnwidth]{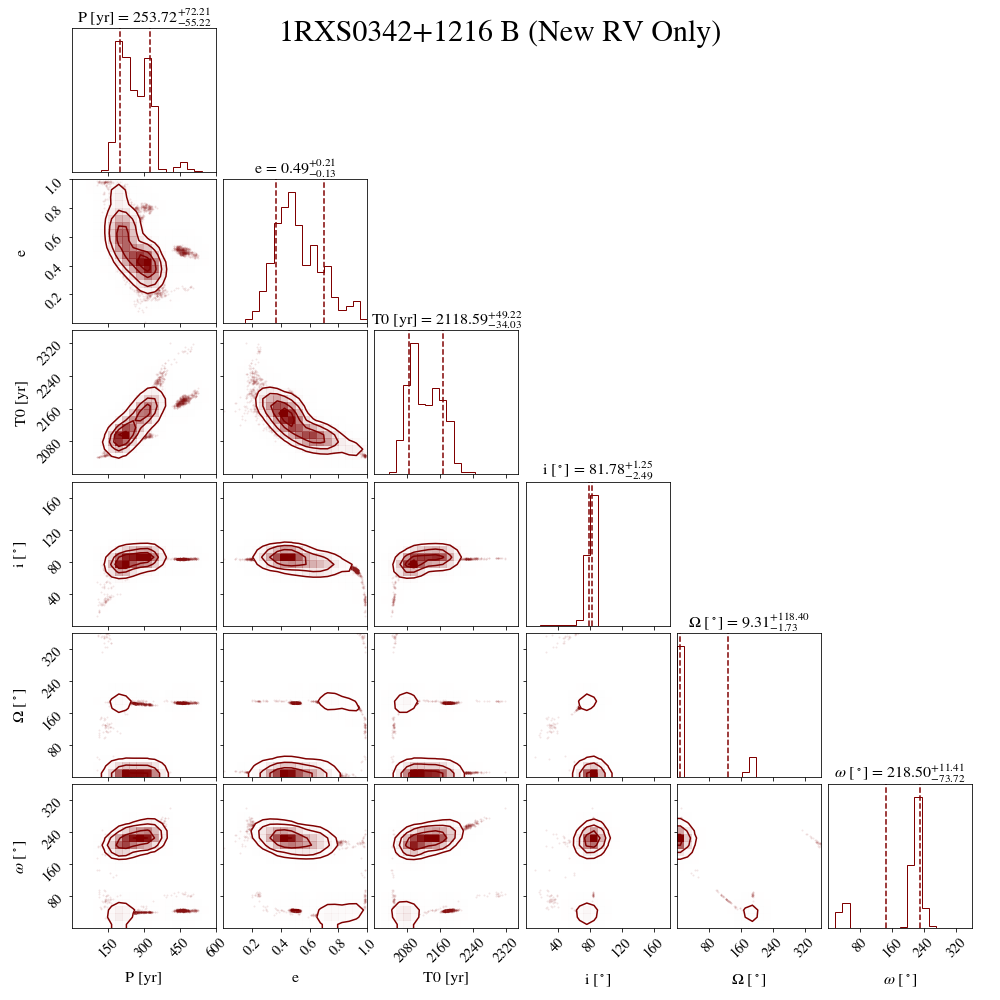}}
\caption{New orbit fit of 1RXS0342+1216 with the 2020 RV data point.  Although RV data has the ability to break the degeneracy in the orientation of the orbital plane, the relative RV between the primary and the secondary in the 2020 epoch was close to 0 km s$^{-1}$, meaning that we are not able to resolve that degeneracy with our data.}
\label{fig:rv.png}
  \end{center}
\end{figure}

 \section{Uniform Priors Fit} \label{uniform}
 In this section, we present our results for the orbit fit of the companion with the traditionally used uniform parameter priors. We split our results into two main subsections: testing system mass changes and the final orbit fit with the NIRC2 astrometry point from 2023A and KPIC RV from 2020 incorporated in the fit. We run the uniform priors with Efit5 as was done for the observable-based prior run as well as with orbitize! (\citealt{Blunt_2020}). The interest in using both methods is because Efit5 uses a nested sampling algorithm, MULTINEST, while orbitize! uses a Markov-Chain Monte Carlo approach. orbitize! Also intakes a Gaussian prior for the system mass, while Efit5 holds the mass fixed at a single value. We aim to verify that both approaches show similar results in the orbital posteriors.

  \subsection{Change in System Mass}
\begin{figure*}[ht]
    \centering
    \centering{{\includegraphics[width= 12cm]{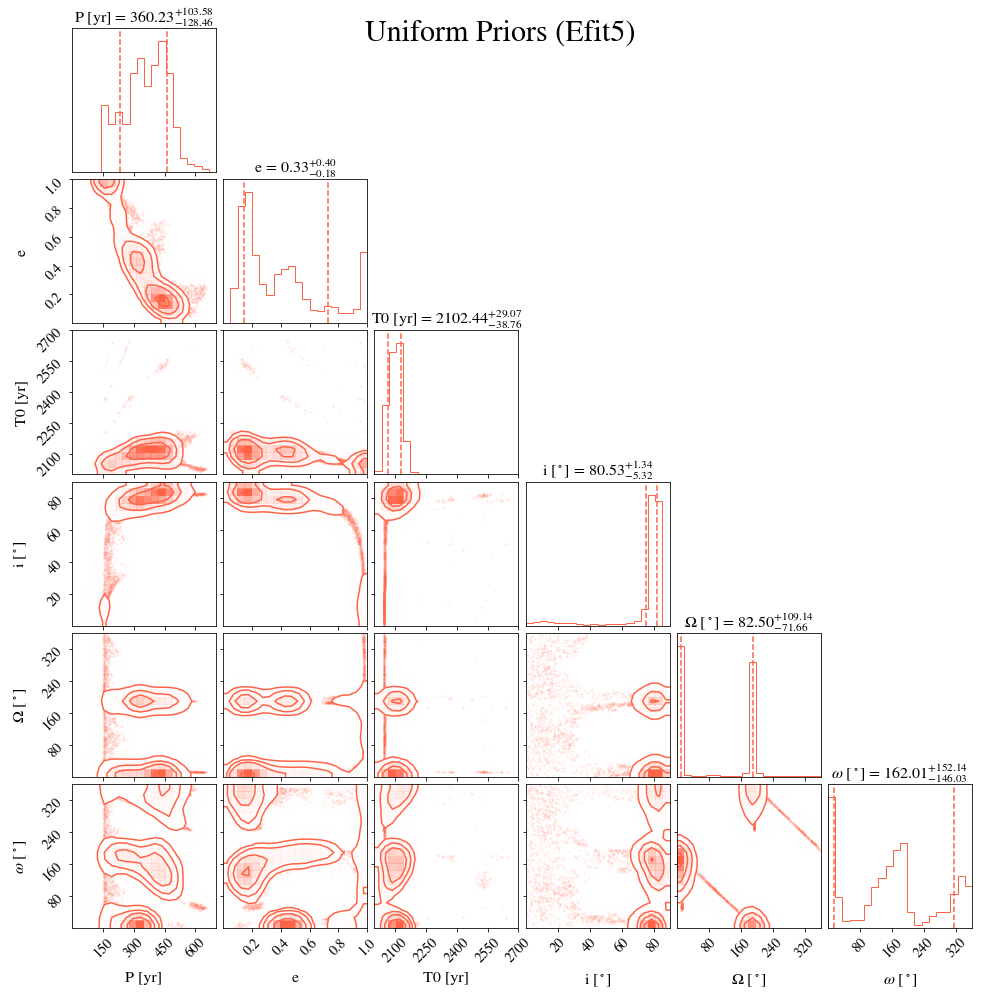} }}%
    \qquad
    \centering{{\includegraphics[width=12cm]{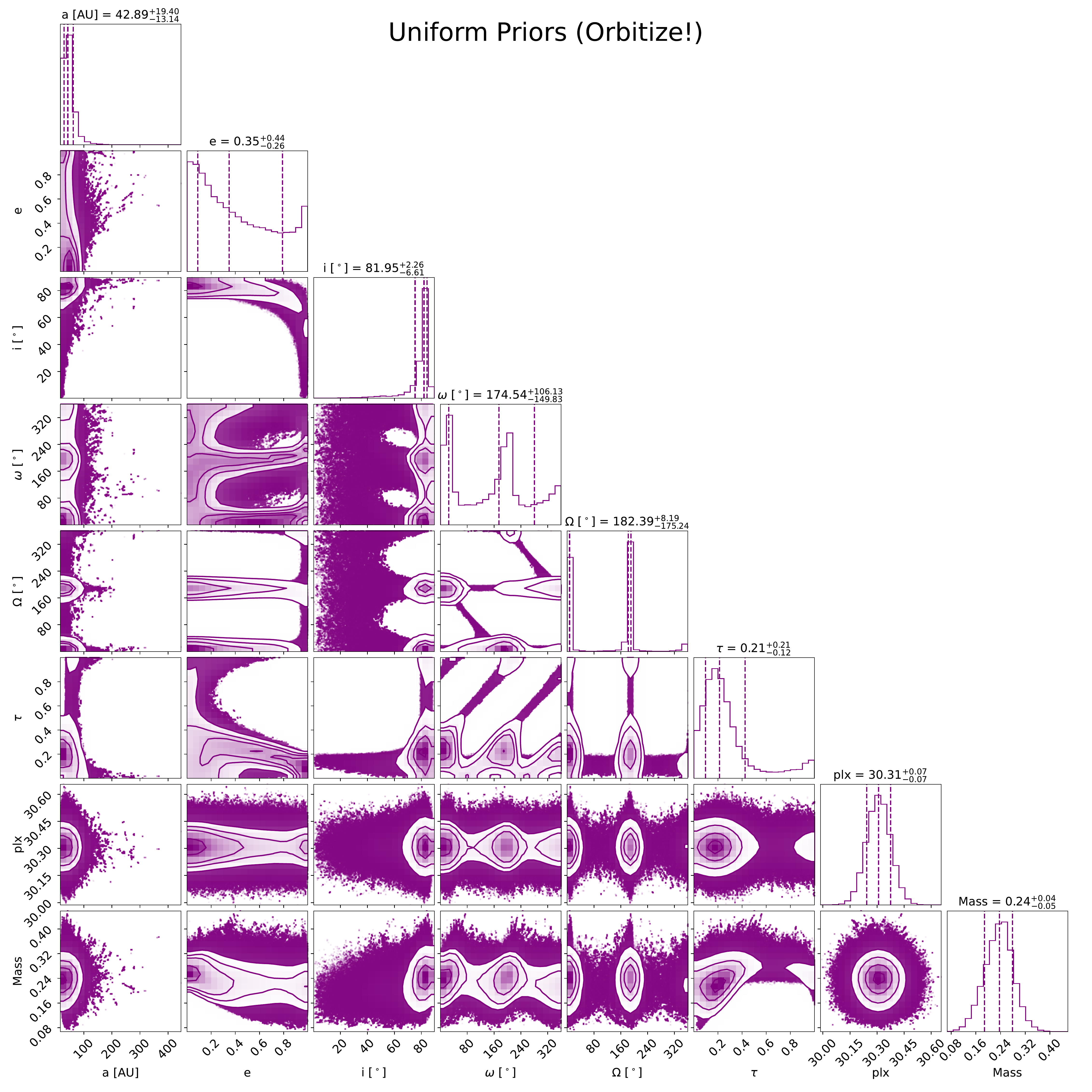} }}%
    \caption{Uniform prior fit with Efit5 and orbitize! for the previous system mass of 0.23 \Msol. }%
    \label{fig:uniform_mass}
\end{figure*}

We run our fits with uniform priors for the old total system mass of 0.23 \Msol. Our results for both the orbitize! and the Efit5 runs are shown in Figure 
\ref{fig:uniform_mass}. We note that in both cases there is a tail of high eccentricities coupled with lower inclinations. In both cases lower eccentricity solutions are also found, which is not the case for the observable prior fit. When the new system mass is incorporated, both tails of high eccentricity are less significant in the orbital fit, with the Efit5 fit's eccentricity tail completely disappearing, similar to the observable prior case.

\begin{figure*}[ht]
    \centering
    \centering{{\includegraphics[width= 12cm]{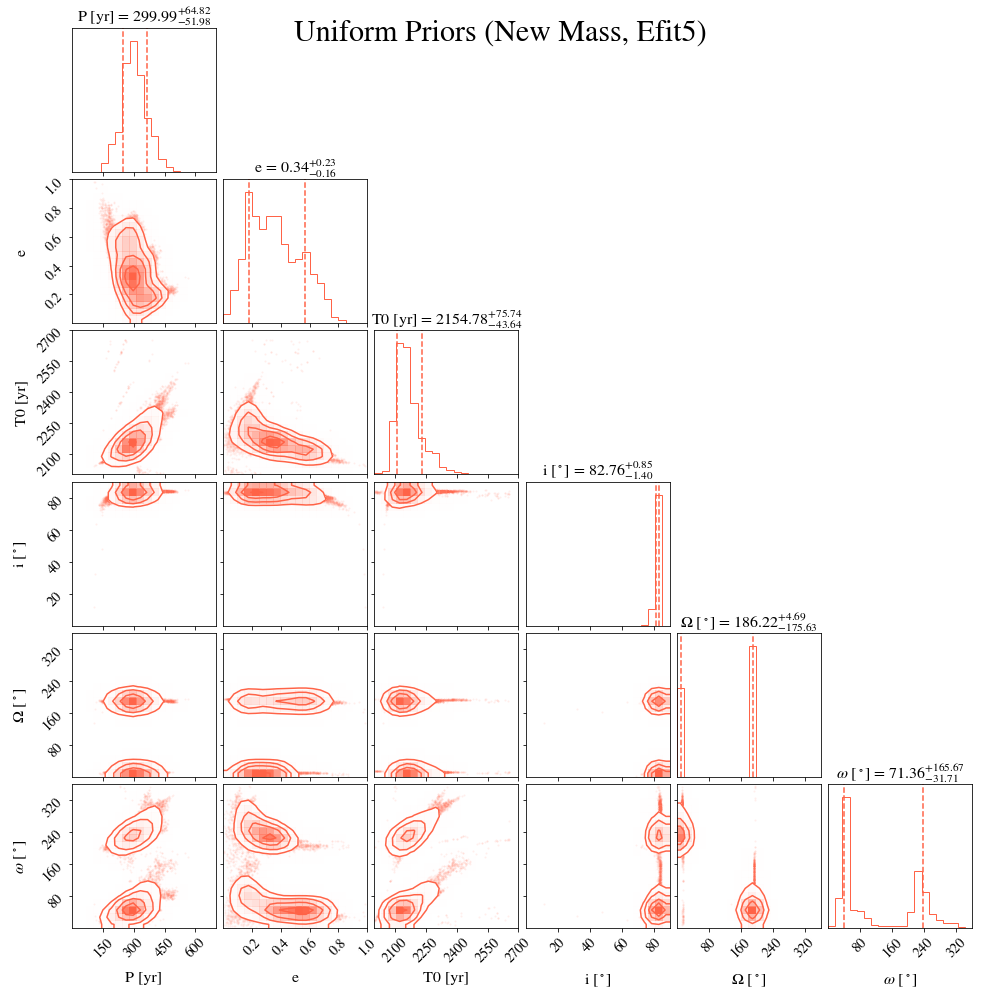} }}%
    \qquad
    \centering{{\includegraphics[width=12cm]{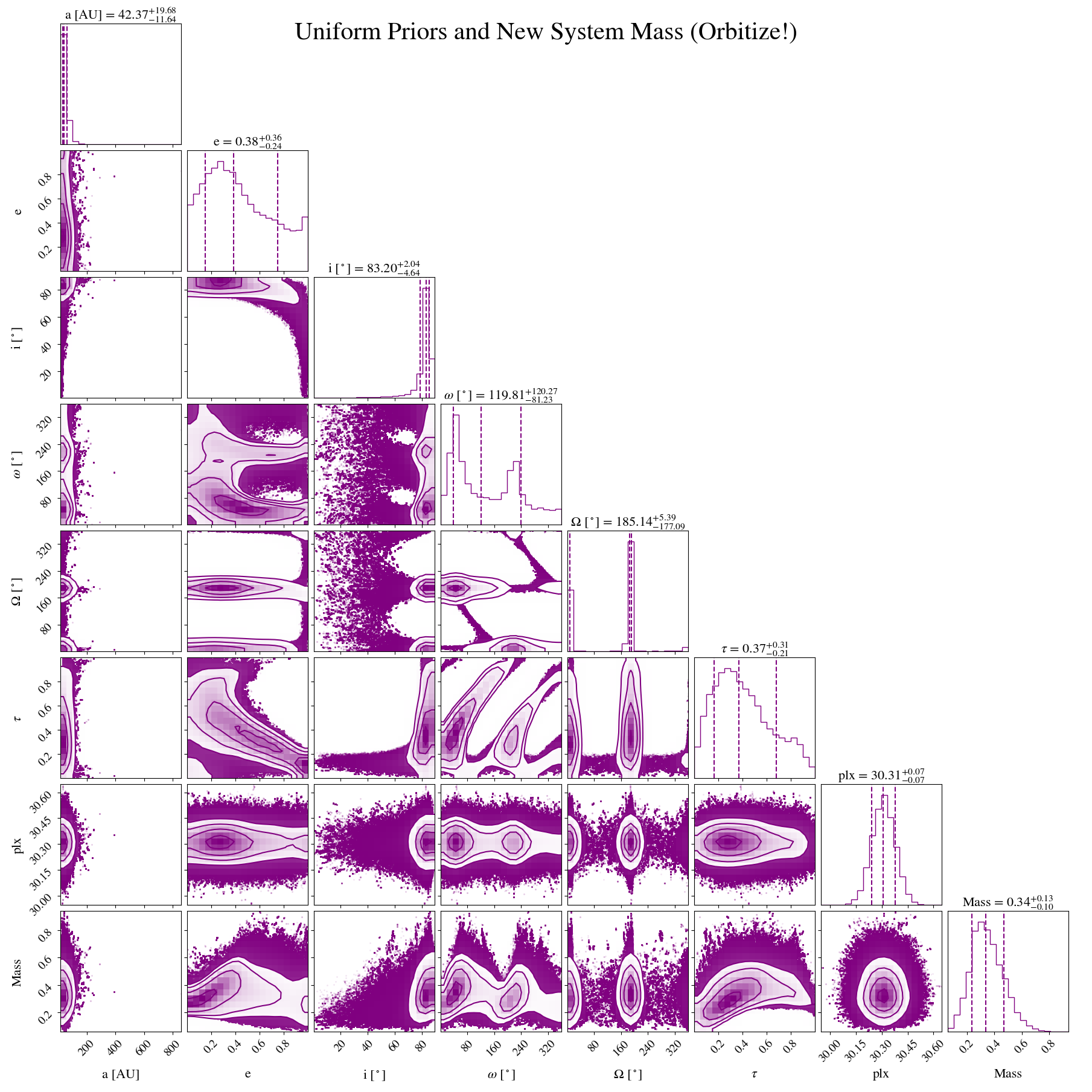} }}%
    \caption{Uniform prior fit with Efit5 and orbitize! for the new system mass of 0.38 \Msol. }%
    \label{fig:uniform_newmass}
\end{figure*}

\subsection{Addition of New Data}

\begin{figure*}[ht]
    \centering
    \centering{{\includegraphics[width= 12cm]{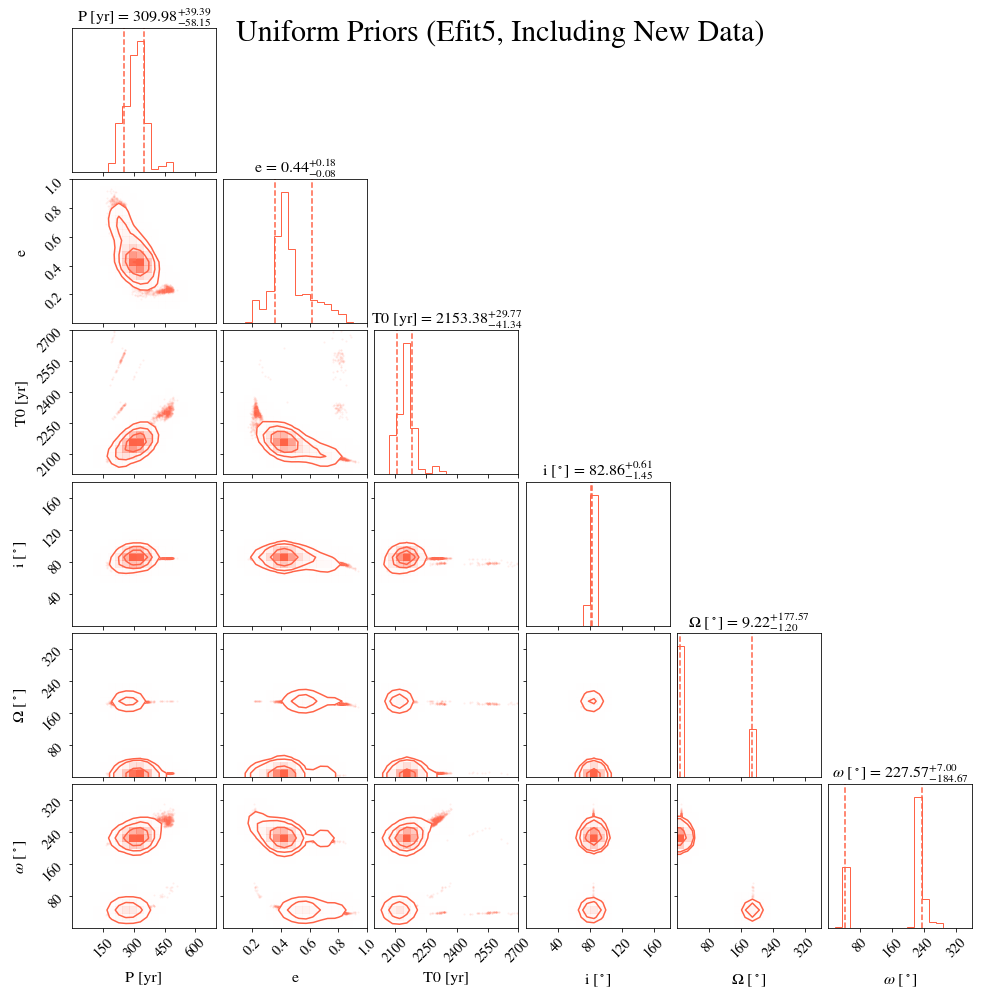} }}%
    \qquad
    \centering{{\includegraphics[width=12cm]{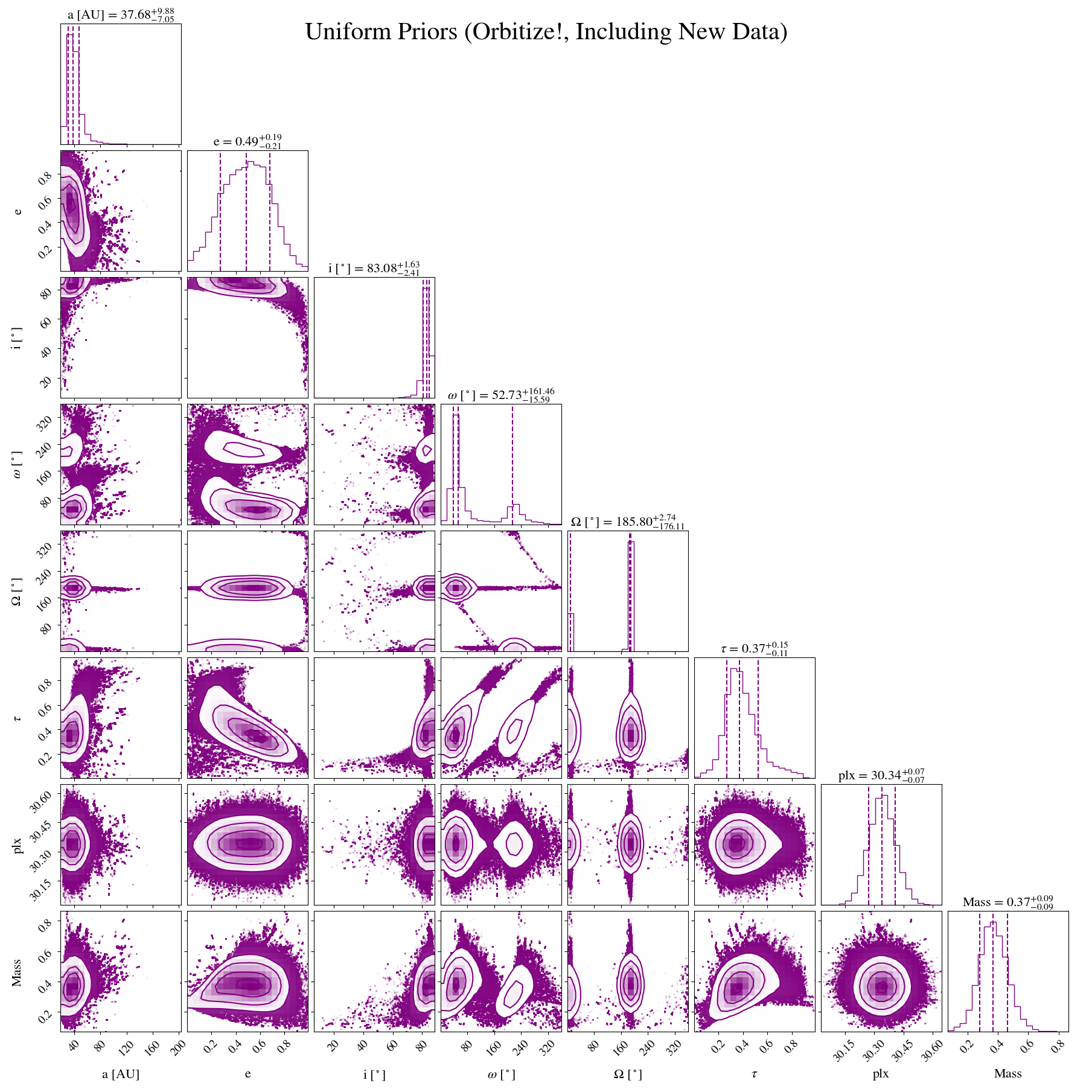} }}%
    \caption{Uniform prior fit with Efit5 and orbitize! for the new system mass of 0.38 \Msol, new 2023A astrometry data point from NIRC2 and 2020 RV data point from KPIC. }%
    \label{fig:uniform_newdata}
\end{figure*}

With the addition of new data, the uniform prior fits show similar results in their posteriors. The corner plot results are shown in 
Figure \ref{fig:uniform_newdata}. In both cases, moderate eccentricities are favored by the data, as is the case for observable priors. The inclinations all agree to a nearly edge-on configuration of i $\sim$ 82 $^{\circ}$.

\section{Mass as Free Parameter}

We also run the orbit fit with observable priors making the mass a free parameter with a uniform prior. We obtain similar posteriors as with a fixed mass for the orbital parameters. The eccentricity is slightly lower than with a fixed mass (0.38 vs. 0.41) and the periastron passage is about 15 years sooner. The mass obtained with models of 0.38 \Msol is well within the uncertainties given by the fit of $0.30\substack{+0.12 \\ -0.06}$ \Msol, which disfavor the previous system mass estimate of 0.23 \Msol.

 \begin{figure}[htb!]
  \begin{center}
\centerline{\includegraphics[width=\columnwidth]{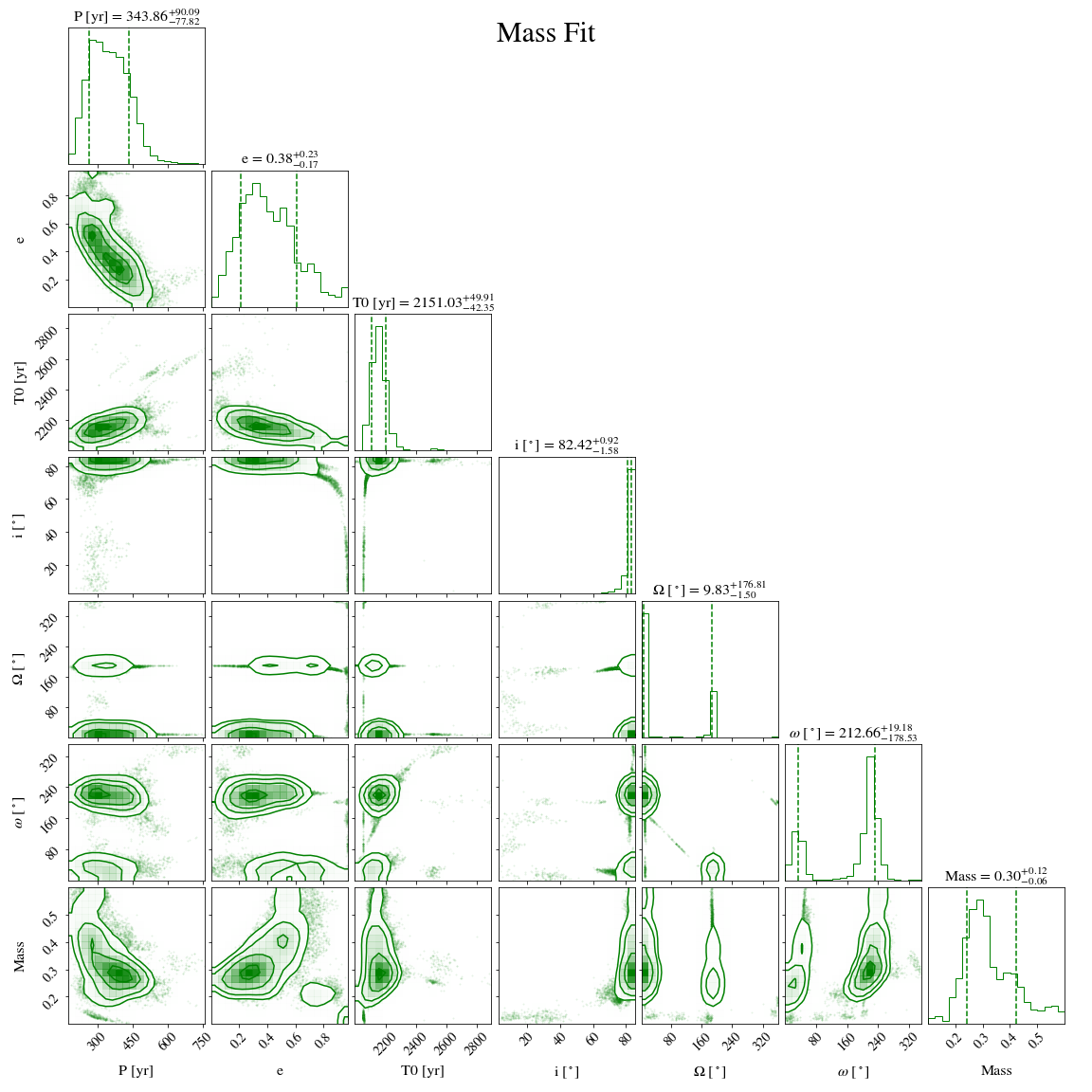}}
\caption{Posteriors for Efit5 with the new data and uniform prior mass fit. Orbital parameters are fit with observable priors.}
\label{fig:mass.png}
  \end{center}

\end{figure}

\end{document}